\documentstyle[12pt]{article}
\newcounter{note}[subsection]
\newcommand{\note}[1]{\stepcounter{note}\footnote{#1}}
\newcommand{\sect}[1]{\section{#1}\setcounter{equation}{0}}

\begin{document}
\renewcommand{\thefootnote}{\fnsymbol{footnote}}
\begin{titlepage}
\begin{flushright}
{\sc Report\ $\sharp$\ \ \ RU94-5-B}\\
\end{flushright}
\vspace{.6cm}
\begin{center}
{\Large{\bf FINITE DEFORMATIONS OF CFT}}\\[5pt]
{\Large{\bf AND SPACETIME GEOMETRY}}\\[30pt]
{\large\it Memory of Victor Nikolaevich Popov}\\[3ex]
{\sc Gregory Pelts
\footnote{Supported by the DOE grant {\em DOE-91ER4651 Task
B}}}\\[2pt]
{\it
Department of Physics\\
The Rockefeller University\\
1230 York Avenue\\
New York, NY 10021-6399}\\[40pt]
{\sc Abstract}\\[12pt]
\parbox{13cm}{
We demonstrate in detail how the space of two-dimensional quantum
field
theories can be parametrized
by off-shell states of free closed string moving in a flat
background.
The dynamic equation corresponding to the condition of
conformal invariance includes an infinite number of higher order
terms,
and we give an explicit procedure for their  calculation.
The symmetries corresponding to
equivalence relations of theories are described.
In this framework
we show how to perform a nonperturbative analysis in
the low-energy limit and prove that it corresponds to the
Brans-Dicke theory of gravity interacting with a skew symmetric
tensor field.}
\end{center}
\end{titlepage}
\renewcommand{\thefootnote}{\arabic{note}}
\tableofcontents
\newpage
\sect{Introduction}
String theory has been formulated in an unusual way.
Usually, physicists start from a description of symmetries,
find a symmetrical classical action,
then quantize it; {\em i.e.}, define the Feynman rules.
In string theory it is quite the opposite.
The Feynman rules (also called Polyakov rules)
are known, and physicists are trying to restore from them
a classical theory and symmetries.
This is important for decoupling of nonphysical
states, understanding the nonperturbative structure
and establishing connection with spacetime geometry.
There is a belief that
classical closed string states can be associated with
quantum conformal field theories in two dimensions (CFT),
which are usually defined
as theories of the single string moving  in some nontrivial
spacetime background. The condition of anomaly cancellation leads to
the
so-called $\beta$-function equation on the background fields.
The main advantage of this approach is its more or less explicit
connection
to spacetime geometry, and the main drawback is that it usually
focuses only on
massless fields.
Treatment of massive fields is problematic,
and, therefore, characterization of dynamical degrees of freedom is
obscure.
Studying  of symmetries is also complicated by the fact that
classically
equivalent CFT may correspond to inequivalent quantum theories.
In addition, it is not clear how to derive the Polyakov diagram
technique
in this approach.

Alternative approaches~\cite{eo,sen,cnw,zw,rsz} are based on the
operator
formalism~\cite{alv}.
In~\cite{eo,sen} a direct connection  has been found  between
(1,1)-primary fields of arbitrary mass level and infinitesimal
deformations of the Virasoro algebra representation.
Later, in~\cite{cnw} it was shown how this work can be related to
deformations
of CFT in the operator formalism.
However, there was a serious problem related to ambiguity of the
vertex
operator commutators, which prevents calculations of deformations
beyond
the first order and of the symmetry algebra structure constants.
The reason for this problem is contact singularities of $T$-products
of
vertex operators inserted close to each other.
In~\cite{me} a correct regularization was outlined for integrals of
such $T$-products.
The condition for  the deformed theory to be
conformally symmetrical induced a dynamic equation for off-shell
vertex operator functions.
This equation is nonlinear
because the regularization is not conformally invariant.
Primary fields represent its solutions in the linear approximation.
Thus, existence of contact singularities makes, in fact, the theory
nontrivial.

{}From our point of view, the ambiguity of vertex operator
commutators is one of
principal and should not be resolved.
It deals with a fact that {\em vertex operators are not exactly
operators},
but some other objects (the definition will be given later in this
paper).
They can be related to some operators. However, such operators act
between
different spaces; {\em i.e.}, are
not automorphisms, and commutators make sense only for automorphic
operators.
In this approach we cannot formulate the string theory in the
language of
some universal algebra of symmetries.
However, this does not make the theory incorrect or less attractive.
Absence of an algebra is
compensated by the structure relating algebraic and geometrical
objects (see
Section~\ref{axioms}).
In fact, what we do is the following. First we define a space $\cal
X$ of $2D$
quantum field theories which includes all equivalence classes and can
be
explicitly parametrized.
Then, symmetries are transformations of  $\cal X$ within
equivalence classes of theories.
Such symmetries do not make a closed algebra and, in fact,  depend on
how we
parametrized the theory or, more exactly, how we defined $\cal X$.
Applying this approach we give an explicit procedure for calculating
all
the higher order terms of the equation of motion, which is a
condition for the
theories to be conformally invariant, and also nonperturbatively
prove that low-energy
deformations of CFT can be described by Brans-Dicke gravity
interacting with
a skew symmetric tensor field. This generalizes the analogous
result~\cite{eg,eg1}
found in the linear approximation.

Many other physicists still believe that a universal string symmetry
algebra
exists.
They are trying to find a {\em right} theory space for which
the symmetry algebra closes. It is too early to evaluate their
results as they
did not proceed far enough to calculate all the commutators and,
therefore,
cannot check whether the symmetry algebra  is closed or not. We
doubt that
it can be done in principle,
at least, without excessive number of variables.
However, some global symmetry groups have been found
(see, for example,~\cite{egn}).

\sect{Axiomatic Conformal Field Theory}
\label{axioms}
Here we will give an axiomatic definition of CFT together with its
motivation and analysis.
In particular, we will show how to introduce vertex operators
and representation of the Virasoro algebra in the suggested axiomatic
framework.
This approach to CFT is similar to the operator formalisms~\cite{alv}
and~\cite{Segal}.

Let us consider a Euclidean (1+1)-quantum
field theory defined on a tubelike world sheet
${\bf S}^1\times{\bf R}^1$. States of the theory
in different moments of time are connected by the propagator
\begin{equation}\label{hm}
\Psi_{t_2}= {\cal S}_{t_2,t_1}\Psi_{t_1},\ \
{\cal S}_{t_2,t_1}=
\exp
\left[
-H( t_2-t_1)
\right]
{}.
\end{equation}
Here $H$ is a Hamiltonian.
{}From a quantum-mechanical point of view, the propagator satisfying
\begin{equation}\label{pgr}
{\cal S}_{t_3,t_1}={\cal S}_{t_3,t_2}{\cal S}_{t_2,t_1}
\end{equation}
completely defines the theory.
However, it is insufficient for world-sheet transformation properties
of the theory.
In order to define such properties we have to assign states  to any
closed oriented contour $\Gamma$ in $\Sigma$ (not necessarily of a
fixed time).
As there is no preferred parametrization for such contours,
the states associated with  differently parametrized contours
should be related by some representation of
the  contour transformation group.
Spaces ${\cal H}^{\Gamma}$ associated with different contours
$\Gamma$ cannot be covariantly identified as they are spaces of
representation of different transformation groups and, therefore,
should be considered independent.
Instead of identifying spaces ${\cal H}^{\Gamma_1}$, ${\cal
H}^{\Gamma_2}$
we should covariantly assign a unitarian operator
between them,
\begin{equation}\label{fr}
{\cal H}^{\Gamma_1}\ \stackrel{\hat{\nu}}{\ \longrightarrow \ }{\cal
H}^{\Gamma_2},
\end{equation}
to each even (keeping orientation) isomorphic map between contours
$\Gamma_1\ \stackrel{\nu}{\ \longrightarrow \ }\Gamma_2$.
It must be done in such a way
that a superposition of two maps $\nu=\nu_1\circ\nu_2$ would
be represented by an operator product
$\hat{\nu}=\hat{\nu_1}\hat{\nu_2}$.
This structure is called a functor.
Thus, we can formulate the following  statement.
\newtheorem{CFT}{Axiom}
\begin{CFT}
There is defined a functor ${\cal H}$ from the category of oriented
closed contours to the category of Hilbert spaces.
\end{CFT}
We can naturally extend this functor to include multicomponent
contours,
which will correspond to multiparticle states.
\begin{CFT}
If the contour $\Gamma$ consists of $N$ components
$\Gamma_1,\ldots,\Gamma_N$,
then
$$
{\cal H}^{\Gamma}=\bigotimes_{i=1}^N{\cal H}^{\Gamma_i}.
 $$
\end{CFT}
Let $\Gamma_{\rm in}$, $\Gamma_{\rm out}$ be two nonintersecting
contours
encircling the world sheet counterclockwise with $\Gamma_{\rm out}$
following
$\Gamma_{\rm in}$ in time. Then, the states in ${\cal H}^{\Gamma_{\rm
out}}$
should be expressed
through the states in ${\cal H}^{\Gamma_{\rm in}}$ by means of a
linear
map generalizing the propagator~(\ref{hm}).
It makes us treat  contours $\Gamma_{\rm in}$, $\Gamma_{\rm out}$
differently,
which violates the covariance of the formalism under time-inverting
transformations.
It can be avoided, if we require the following:
\begin{CFT}\label{CP}
The spaces corresponding to the contrary-oriented contours  are
conjugated.
\end{CFT}
Then the propagator can be considered as an element of ${\cal
H}^{\partial\Sigma}$,
where $\Sigma$ is part of the world sheet enclosed between the
contours.

Now let the world sheet be endowed with a conformal structure.
For oriented two-dimensional surfaces it is the same as a complex
structure.
The propagator of a conformally symmetrical theory must be
conformally invariant.
Therefore, to  define a propagator, we only need to know
the conformal structure of the part of the word tube  corresponding
to the process and
do not need to refer to its particular position in the tubelike
world-sheet.
Moreover, using the sewing procedure
(see Axiom \ref{sew} below) we can assign an element of
${\cal H}^{\partial\Sigma}$ to
Riemann surfaces which cannot be mapped to the tube. They may have
an arbitrary number of handles  and boundary components.
If the boundary of $\Sigma$ consists of more then two components,
it will be a scattering amplitude with the number of handles
equal to the number of the diagram loops.
A Riemann surface $D$ of the disk topology
(no handles, one boundary component) can be considered as the
compactified
{\em past} (the part of the world-sheet consisting of points with
$t \leq t_0$) and we can assign to it a vacuum state.
Altogether, it can be formulated in  the following way.
\begin{CFT}
To any bordered Riemann surface $\Sigma$ there corresponds in a
conformal
invariant way a specific element
$\langle 0 \rangle_{\Sigma}$  of ${\cal H}^{\partial\Sigma}$.
\end{CFT}
A multicomponent surface represents a set of
independent scattering processes. The corresponding amplitude is a
product
of amplitudes of these processes.
\begin{CFT}
If the surface $\Sigma$ consists of $N$ components
$\Sigma_1\ldots\Sigma_N$, then
$$
{\langle 0 \rangle}_{\Sigma}=\bigotimes_{i=1}^N
{\langle 0 \rangle}_{\Sigma_i}.
 $$
\end{CFT}
To the anticonformal bijection between Riemann surfaces there
corresponds
an odd map between their boundaries and, therefore, according to
Axiom~\ref{CP}, an antilinear operator between associated spaces.
For a CP-invariant theory, the amplitude must be invariant under
such antilinear operators. This requirement can be formulated as
follows.
\begin{CFT}\label{real}
Amplitudes corresponding to Riemann surfaces with conjugated complex
structure are conjugated to each other.
\end{CFT}
The propagator must establish a transitive
relation  between state spaces associated with different moments of
time,
and the amplitudes must  be compatible with this relation.
This property can be covariantly formulated as follows.
\begin{CFT}[sewing]\label{sew}
The amplitude corresponding to the surface $\Sigma$ can be expressed
through the amplitude corresponding to the surface $\Sigma_{\Gamma}$,
resulting
from the
cutting $\Sigma$ along  the closed contour $\Gamma$,  by the formula
\begin{equation}\label{sewing}
\langle 0 \rangle_{\Sigma} = Sp_{\Gamma}
{\langle 0 \rangle}_{\Sigma_{\Gamma}} .
\end{equation}
Here $Sp_{\Gamma}$ is an operator contracting components of ${\cal
H}^{\partial\Sigma_{ \Gamma}}$
corresponding
to two copies of the contour $\Gamma$ with opposite orientations.
\end{CFT}
This axiom generalizes the property~(\ref{pgr}) of a  conventional
quantum
mechanical  propagator.

The set of axioms above can be thought of as a definition of CFT.
We will show that all the important attributes of CFT, such as
vertex operators and a representation of Virasoro algebra,
can be derived from them.
In order to describe a CFT with a central charge we should relax the
axioms above,
changing them to their projective analogues.
Then, instead of the functor
to the category of Hilbert spaces, we will
have a functor to the category
of projective Hilbert spaces, and amplitudes, vacuum spaces and
propagators
will be defined only up to a constant multiplier.

For a chirally symmetrical CFT the spaces corresponding to
contrary-oriented
contours can be identified.
Therefore, instead of the functor from the
category of oriented contours,
we will have a functor from the category of
nonoriented contours; {\em i.e.},
maps changing contour orientation will be  also represented.
Then, according to axiom~\ref{CP}, spaces ${\cal H}^{\Gamma}$ must be
selfconjugated; {\em i.e.}
they  must be  endowed with antilinear automorphism $C$,  satisfying
$$
C^2=\chi,\hspace{1em}
(C\eta,\xi)=\chi(C\xi,\eta)\hspace{2em}
\left( \chi=\pm 1
\right).
 $$
Here delimiters $(*,*)$ denote a Hilbert product.
The form $(c*,*)$ is bilinear. It is symmetric if $\chi=1$ and
skew symmetric
otherwise.
The {\em skew symmetric} option is available only in the case
of nontrivial central charge. It can be shown that in the
chiral symmetrical case, an anomaly can be ruled out from the functor
which can be defined in nonprojective way.
However, a constant multiplier still may appear in~(\ref{sewing}),
which
may induce nontrivial central charge in this case.

\subsection{Vertex operators}
We will say that an element
${\langle\Psi\rangle}_{\Sigma}$ of ${\cal H}^{\partial\Sigma}$  has
{\em support} in a closed set $S\subset\Sigma$,
if for any contour $\Gamma$ surrounding $S$ counterclockwise
(or a set of contours if $S$ consists of more then one component)
it can be represented as
$$
{\langle\Psi\rangle}_{\Sigma}=Sp_{\Gamma}{\langle 0
\rangle}_{\Sigma_{\rm ext}}\otimes
\langle{\Psi\rangle}_{\Sigma_{\rm in}}.
 $$
Here $\Sigma_{\rm in}$ and $\Sigma_{\rm ext}$  are,
respectively, the internal and external parts of $\Sigma$ divided by
the
contour $\Gamma$, and
$\langle\Psi\rangle_{\Sigma_{\rm in}}$
is an element of ${\cal H}^{\Gamma}$.
Considering the map ${\langle\Psi\rangle}_{\Sigma}\rightarrow
\langle\Psi\rangle_{\Sigma_{\rm in}}$ as an
equivalence
relation, we can
identify the states corresponding to different areas of the
Riemann surface but having the same support.
Let us denote the space resulting from such identification as $H_S$.
For its elements we will use the capital Greek letter and for their
images in
${\cal H}^{\partial\Sigma}$ the same letters enclosed in angular
brackets with a superscript
index referring to the related Riemann surface; for example,
\begin{equation}\label{embed}
\Psi\ \longrightarrow \ {\langle\Psi\rangle}_{\Sigma}\hspace{1em}
\left(
\Psi\in H_S,\hspace{1em}{\langle\Psi\rangle}_{\Sigma}\in{\cal
H}^{\partial\Sigma}
\right).
\end{equation}
To endow $H_{S}$ with topology, we will call a sequence
in $H_{S}$ vanishing, if for any $\Sigma_{\rm in}$
such as
$\Sigma_{\rm in}\setminus\partial\Sigma_{\rm in}\supset S$
it projects to a vanishing sequence in ${\cal H}^{\partial\Sigma_{
\rm in}}$.
It is easy to see that $H_{S}$  then will be  a Banach space.
Normally, images of propagators and, therefore, images of projections
are dense in ${\cal H}^{\partial\Sigma}$. Thus, the difference
between the spaces $H_S$ and
${\cal H}^{\partial\Sigma}$
is, in some sense, topological.
As we will see later in this section,
states with support in a single point can be interpreted as vertex
operators.
More exactly, it can be thought of as an exact definition of what is
usually
called vertex operators.
The spaces  $H_z\equiv H_{\{z\}}$ $(z\in\Sigma)$
form a bundle on $\Sigma$,
sections of which we will call vertex operator functions.
To show that elements of this spaces are, indeed, vertex operators,
we should first define their $T$-product.
It can be done as follows:
\begin{equation}
\langle\Psi_0\cdots\Psi_n\rangle_{\Sigma}
=Sp_{\partial\Sigma_{\rm ext}}
\bigotimes_{i=0}^n
{\langle\Psi_i\rangle}_{D_i}\otimes{\langle 0 \rangle}_{\Sigma_{\rm
ext}}
\hspace{1em}
\left(
\Psi_i\in H_{z_i}
\right) .
\label{t-p}\end{equation}
Here $D_i\hspace{1em}(i=0,N)$ are subsets of $\Sigma$
such as
$$
z_i\in D_i,\hspace{1em} D_i\bigcap D_j=\emptyset\hspace{1em}(i\neq
j),
 $$
and $\Sigma_{\rm ext}$ is their complement in $\Sigma$.
This $T$-product is a multilinear map from
$H_{z_0}\times\cdots\times H_{z_N}$ to ${\cal H}^{\partial\Sigma}$.
Acting analogously to~(\ref{t-p}), we can put into correspondence to
element of $H_z$ operators from $H_S$ to $H_{S^{\prime}}$ if
$z\in\left(
S^{\prime}\setminus\partial S^{\prime}
\right)
\setminus S$.
Note that $S\subset S^{\prime}$. Therefore, such operators are
never automorphisms, and their commutators should not be defined.

The Virasoro algebra does not have a bounded natural representation
in
${\cal H}^{\partial\Sigma_{ \rm in}}$. The conformal transformations
deform the boundary of
$\Sigma_{\rm in}$ and, therefore,
corresponding to them linear operators are not
automorphisms,
as they act between
Hilbert spaces associated with different contours.
However, we can define such a representation
in the space $H_{z}$, which is independent of the position
of the boundary.

\subsection{Local multipliers}
Let $\Gamma$ be a closed contour, which divides  $\Sigma$ into
$\Sigma_1$ and $\Sigma_2$.
We will say that an operator ${\cal O}$ in ${\cal H}^{\Gamma}$ has
support in
$S\subset\Gamma$,
if the state $Sp_{\Gamma}{\langle 0 \rangle}_{\Sigma_2}\otimes{\cal
O}{\langle 0 \rangle}_{\Sigma_1}$
has support in $S$, and
denote the algebra of such operators as
${\cal L}_{\Gamma,S}$.
We will call  an invertible operator $U:\ {\cal H}^{\Gamma}\
\longrightarrow \ {\cal H}^{\Gamma}$ a
{\em local multiplier},
if a similarity transformation generated by $U$ does not enlarge the
support of any operator;  {\em i.e.\/},
\begin{equation}\label{lm}
\forall S\in\Gamma:\hspace{1em}U{\cal L}_{\Gamma,S}U^{-1}
\subseteq{\cal L}_{\Gamma,S}.
\end{equation}
It can be shown that the group  of local multiplier does not depend
on how
the contour is placed on the Riemann surface
and is not affected by CFT deformations.
It makes this group an important tool in the description
of deformed theories and symmetries between them.

\subsection{Residuelike operations}
We will call a linear operator  from
the space of functions on
$\Sigma^{N+1}\setminus\Delta
\!\left(
\Sigma^{N+1}
\right)$
\note{The symbol
$\Delta\!\left(
\Sigma^{N+1}
\right)$
denotes diagonal subset
of Dekart product $\Sigma^{N+1}$, elements of which contain at least
one pair of equal points.}
to the space of functions on $\Sigma$ a residuelike operator
of rank N,
if the value of its image function at the point $z$ in $\Sigma$
is determined by the behavior of its argument function in an
arbitrarily
small environment of the  point in $\Sigma^{N+1}$ with all the
coordinates
equal to $z$. We will denote such operations in one of the following
ways
\begin{eqnarray}
G(z_0)&=&{\cal R}_{z_N=\cdots=z_0}F(z_1,\ldots,z_0)\nonumber\\
&&\hspace{1.5em} \mbox{ or }\nonumber\\
G(z_0)&=&{\cal R}_{\bar{z}_N=\cdots=\bar{z}_0}F(z_1,\ldots,z_0).
\nonumber\end{eqnarray}
The zero rank residuelike operations
are simply local operators in the space of functions,
for example, differential operators.
The conventional residue is an example of a rank $1$
residuelike operation defined for holomorphic functions.

Using $T$-product~(\ref{t-p}) we can define representation of
residuelike operations by
multilinear products in the space  of vertex operator functions:
\begin{eqnarray}
\{\Psi_i\}_{i=1}^N\ \longrightarrow \ \
\Upsilon&=&{\cal R}_{z_N=\cdots=z_0}\Psi_0\cdots\Psi_N,\nonumber\\
\langle\Upsilon(z_0)\rangle_{\Sigma}
&\stackrel{\rm def}{=}&
{\cal R}_{z_N=\cdots=z_0}
\langle\Psi_0\cdots\Psi_N\rangle_{\Sigma}.
\label{r-l}\end{eqnarray}
This representation applied to the differential
operators\note{
The symbols $\partial_{z}$, $\partial_{\bar{z}}$ denote holomorphic
and antiholomorphic
derivatives. If the variable of differentiation should not be
specified,
we will also denote them as $\partial$, $\bar{\partial}$.}
$\partial_{z}$, $\partial_{\bar{z}}$
produces a flat nonintegrable connection for the vertex operator
bundle.

\subsection{Energy-momentum tensor}
An infinitesimal deformation of the conformal structure can be given
as
$$
\delta(d\!z)=\varepsilon_{\bar{z}}^{z}d\!\bar{z},\hspace{1em}
\delta(d\!\bar{z})=\varepsilon_{z}^{\bar{z}}d\!z,
 $$
where $\varepsilon_{\bar{z}}^{z}$, $\varepsilon_{z}^{\bar{z}}$
are so-called Beltrami differentials. Applying the sewing
property (axiom~\ref{sew}), one can show that infinitesimal
deformation of the amplitudes
are additive, {\em i.e\/},
$$
\delta\!{\langle 0 \rangle}_{\Sigma} =
\frac{1}{\pi}\int_{\Sigma}
(\varepsilon_{\bar{z}}^{z}\langle T_{zz}\rangle_{\Sigma}
+\varepsilon_{z}^{\bar{z}}
\langle T_{\bar{z}\bar{z}}\rangle_{\Sigma})
\,d^2\!z .
 $$
Here $T_{zz}$, $T_{\bar{z}\bar{z}}$ are some vertex operator
functions.
The  Beltrami differentials
$$
\varepsilon_{\bar{z}}^{z}=\partial_{\bar{z}}v^z,\hspace{1em}
\varepsilon_{z}^{\bar{z}}=\partial_z v^{\bar{z}}
 $$
can be produced by the infinitesimal world sheet transformation
$$
\delta\!z = v^z,\hspace{1em} \delta\!\bar{z}=v^{\bar{z}}.
 $$
Therefore, the corresponding deformation of amplitude must be
trivial;
{\em i.e.},
\begin{eqnarray}
0&=&\frac{1}{\pi}\int_{\Sigma}
\left(\partial_{\bar{z}}v^z
\langle T_{zz}\rangle_{\Sigma}
+\partial_z v^{\bar{z}}
\langle T_{\bar{z}\bar{z}}\rangle_{\Sigma}
\right)
\,d^2\!z
\nonumber\\[1ex]
&=&
-\frac{1}{\pi}\int_{\Sigma}
\left(
v^z \partial_{\bar{z}}
\langle T_{zz}\rangle_{\Sigma}
+v^{\bar{z}}\partial_z
\langle
T_{\bar{z}\bar{z}}
\rangle_{\Sigma}
\right)
\,d^2\!z
\nonumber\\[1ex]
&=&-
\frac{1}{\pi}\int_{\Sigma}
\left(
v^z\langle
\partial_{\bar{z}}T_{zz}
\rangle_{\Sigma}
+v^{\bar{z}}\langle
\partial_z T_{\bar{z}\bar{z}}
\rangle_{\Sigma}
\right)
\,d^2\!z .
\nonumber\end{eqnarray}
This condition will be satisfied for any $v^z$, $v^{\bar{z}}$ only if
\begin{equation}\label{hol}
\partial_{\bar{z}} T_{zz}=\partial_{z} T_{\bar{z}\bar{z}}=0.
\end{equation}
It can be shown that $T_{zz}$, $T_{\bar{z}\bar{z}}$
are left and right components of the conformal
energy-momentum tensor;  {\em i.e.\/},
the representation of the left and right Virasoro algebra
in $H_{z_0}$ can be expressed through them
\begin{eqnarray}
L_{v^z}\Upsilon(z_0)&=&
{\rm Res}_{z=z_0}v^z T_{zz}(z)\Upsilon(z_0),
\nonumber\\[1ex]
\bar{L}_{v^{\bar{z}}}\Upsilon(z_0)&=&
{\rm Res}_{\bar{z}=\bar{z}_0}
v^{\bar{z}}T_{\bar{z}\bar{z}}(z)\Upsilon(z_0).
\label{rep}\end{eqnarray}
Here $v^z$ and $v^{\bar{z}}$ are, respectively, holomorphic and
antiholomorphic
tangent fields, and $L_{v^z}$ and $\bar{L}_{v^{\bar{z}}}$ are
the corresponding
generators of the representation.
For generators corresponding to the basis
elements
of the Virasoro algebra
$$
v_n^z=(z-z_0)^{n+1},\hspace{1em}
v_n^{\bar{z}}=(\bar{z}-\bar{z}_0)^{n+1}\hspace{1em}
\left(
n\in{\bf Z}
\right)
 $$
we will use the canonical notations
$$
L_n\equiv L_{v_n^z},\hspace{1em}
\bar{L}_{n}\equiv\bar{L}_{v_n^{\bar{z}}}.
 $$

\sect{Infinitesimal Deformation of CFT}
As is known,
the closed string states can be described
as elements of the space
of the Virasoro representation $H_{z_0}$
satisfying the set of equations
\begin{equation}
(L_k + \delta_{k,1})\Psi(z_0)=
(\bar{L}_k +\delta_{k,0})\Psi(z_0) =0\hspace{1em} (k\geq0).
\label{canon}\end{equation}
Using parallel transportation $z\rightarrow z+c$ we can transform
$\Psi(z_0)\in H_{z_0}$ to the state $\Psi(z_0+c)\in H_{z_0+c}$
satisfying the analogous set of equations.
In this way, we can
define $\Psi$ as a translationally invariant vertex operator
function.
The formula~(\ref{canon}) together with the condition of  translation
invariance
\begin{equation}\label{transl}
(d+L_{-1})\Psi=( \bar{d}+\bar{L}_{-1})\Psi=0
\end{equation}
can be written as
\begin{equation}
L_{v^z}\Psi+
\partial_{z} v^z\Psi=\bar{L}_{v^{\bar{z}}}\Psi+
\partial_{\bar{z}} v^{\bar{z}}\Psi=0
\hspace{1em}
(\partial_{\bar{z}}v^z=\partial_z v^{\bar{z}}=0).
\label{ci}\end{equation}
This  means that $\Psi$ is  a (1,1)-primary field;  {\em i.e.\/},
a conformally invariant vertex operator
function of the conformal dimension $(1,1)$.
Such primary fields
parametrize infinitesimal deformations of CFT:
\begin{equation}\label{infdef}
{\langle 0 \rangle}_{\Sigma}\ \longrightarrow \ {\langle 0
\rangle}_{\Sigma} +\delta{\langle 0 \rangle}_{\Sigma},\hspace{1em}
\delta{\langle 0 \rangle}_{\Sigma}=
\frac{1}{\pi}\int_{\Sigma}
{\langle\Psi\rangle}_{\Sigma}
\,d^2\!z.
\end{equation}
The conformal invariance of the amplitudes deformed in this way
will be retained if we deform the Virasoro representation
as
\begin{eqnarray}
L_{v^z}\ \longrightarrow \  L_{v^z} +\delta\!L_{v^z},\hspace{1em}
\delta\!L_{v^z} &=& \frac{1}{2\pi i}\oint_{\Gamma}\Psi
v^z\,d\!{\bar z};\nonumber\\[1ex]
\bar{L}_{v^{\bar{z}}}\ \longrightarrow \
\bar{L}_{v^{\bar{z}}}
+\delta\!\bar{L}_{v^{\bar{z}}},\hspace{1em}
\delta\!\bar{L}_{v^{\bar{z}}} &=&
\frac{1}{2\pi i}\oint_{\Gamma}
\Psi v^{\bar{z}}
\,d\!z.
\label{mark}\end{eqnarray}
This is equivalent to the formula for infinitesimal
deformations of the Virasoro representation
proposed in~\cite{eo,sen}.

\subsection{Deformation of vertex operators}
In order to understand how to proceed deformations beyond the first
order we must describe the space of vertex operators of
infinitesimally
deformed theory.
Let us identify spaces of the vertex
operators of the infinitesimally deformed and initial theories.
In order to do it, we should deform the map~(\ref{embed})
making its image consist of the states having point support with
respect to the deformed propagator~(\ref{infdef}).
Formally, it can be done as follows:
\begin{equation}
\langle\Upsilon\rangle_{\Sigma}
\ \longrightarrow \
\langle\Upsilon\rangle_{\Sigma}
+ \delta\!\langle\Upsilon\rangle_{\Sigma}
,\hspace{1em}
\delta\!
\langle\Upsilon(z_0)\rangle_{\Sigma}=
\frac{1}{\pi}\int_{\Sigma}
\langle\Psi(z)\Upsilon(z_0)\rangle_{\Sigma}
\,d^2\!z .
\label{infvrt}\end{equation}
However, in general,
the integral on the right-hand side of this formula
may be divergent because of the contact singularity of the
$T$-product.
This reflects the absence of conformally invariant connection for the
vertex
operator bundle on the theory space.
Otherwise, all deformed theories would be equivalent to the initial
one.
If we use in~(\ref{infvrt}) the simple cutoff regularization
\begin{eqnarray}
\delta_R\langle\Upsilon(z_0)\rangle_{\Sigma}=
\frac{1}{\pi}\int_{\Sigma\setminus D_{z_0,R}}
\langle\Psi(z)\Upsilon(z_0)
\rangle_{\Sigma}
\,d^2\!z\nonumber\\[1ex]
\left(
D_{z_0,R}=\left\{z\in\Sigma,\, |z-z_0|\leq R\right\}
\right),
\nonumber\end{eqnarray}
the deformed vertex operators will have support in the disk
$D_{z_0,R}$
rather then in the single point $z_0$.
States with support in the disk can be also produced by an average of
the cutoff regularizations with smaller radii
\begin{equation}
\delta\!\langle
\Upsilon(z_0)
\rangle_{\Sigma}=
\int_0^{\infty}\delta_r
\langle\Upsilon(z_0)
\rangle_{\Sigma}d\!\mu(r).
\label{reg}\end{equation}
Here $d\!\mu$ is a generalized measure in ${\bf R}_+$ having support
in
$[0,R]$ and integrable in a  product with all the functions having a
finite degree singularity at $r=0$.
Diminishing support of the measure we can diminish
the support of the state. If we use the measure with
support in zero, states will have one-point support.
Such a measure exists  and can be defined by the formula
\begin{equation}\label{measure}
\int_0^{\infty} r^{2\alpha} \mu (r)\,d\!r =\Lambda(\alpha)\ \
(\alpha\in{\bf R}).
\end{equation}
Here, $\Lambda$ is a smooth function on ${\bf R}$ satisfying
\begin{equation}\label{bound}
\Lambda(0)=1,\hspace{1em}\Lambda(\alpha)=0\hspace{1em}
(\alpha>A)
\end{equation}
for some positive $A$.
A related proposal for regularization corresponding to specific
stepfunction $\Lambda=\Theta(-\alpha)d^{\alpha}$ was independently
made
in~\cite{rsz}. Discontinuity of this $\Lambda$ at $r=0$ creates
ambiguity for
calculation, especially in the low-energy limit.
For convenience we will impose on $\Lambda$  the condition
$$
\Lambda(k)=0\hspace{1em}(k\in{\bf Z}, k\geq 1).
 $$
Then the regularization will not affect integrals of regular
functions.
The deformation~(\ref{infvrt}) of embedding~(\ref{embed}) induces a
analogous deformation of the $T$-product~(\ref{t-p}):
\begin{equation}
\delta\!
\langle
\Upsilon_0(z_0)\cdots\Upsilon_n(z_n)
\rangle_{\Sigma}
=\frac{1}{\pi}\int_{\Sigma}
\langle
\Psi(z)\Upsilon_0(z_0)\cdots\Upsilon_n(z_n)
\rangle_{\Sigma}
\,d^2\!z.
\label{inftp}\end{equation}
Here the regularization~(\ref{reg}) must be applied separately for
each
of
the vertexes.

\subsection{Deformation of representation of residuelike operations}
\label{drlo}
The action~(\ref{r-l}) of residuelike operations on vertex operator
functions
depends on the $T$-product and, therefore, should be deformed
together with it.
For the infinitesimal deformation of the $T$-product~(\ref{inftp}),
the corresponding deformation of residuelike operations will be
\begin{equation}
\delta\!{\cal R}_{z_n=\cdots=z_0}
\Upsilon_0(z_0)\cdots\Upsilon_n(z_n)
={\cal R}_{z\doteq z_n=\cdots=z_0}
\Upsilon_0(z_0)\cdots\Upsilon_n(z_n)\Psi(z).
\label{rd}\end{equation}
Here ${\cal R}_{z\doteq z_n=\cdots=z_0}$ is
a next rank residuelike operation
defined as
\begin{equation}
{\cal R}_{z\doteq z_n=\cdots=z_0}F={\cal R}_{z_n=\cdots=z_0}
\frac{1}{\pi}\int_{\Sigma}F\,d^2\!z
-\frac{1}{\pi}\int_{\Sigma}
{\cal R}_{z_n=\cdots=z_0}F
\,d^2\!z  .
\label{sc}\end{equation}
It is, indeed, a residuelike operation, because  the right-hand side
of~(\ref{sc})
does not depend on the area of integration as far as it includes
$z_0$.
We will call this operation a successor of ${\cal
R}_{z_{z_n=\cdots=z_0}}$.

The residue theorem can be generalized for contour integrals of
nonholomorphic functions having a finite number of point
singularities
\begin{equation}\label{rt}
\frac{1}{2\pi i}\oint_{\partial\Sigma}G
\,d\!z
=\frac{1}{\pi}\int_{\Sigma}
\partial_{\bar{z}} G
\,d^2\!z
+\sum_{i=1}^n{\rm Res}_{z= z_i}G.
\end{equation}
Here ${\rm Res}_{z= z_i}$ is a generalized residue operation defined
for nonholomorphic functions as an average of contour integrals
around
the circles $C_{z_i,r}=\partial D_{z_i,r}$ with the
measure~(\ref{measure})
\begin{equation}\label{drlr}
{\rm Res}_{z=z_i}F\stackrel{\rm def}{=}\int_0^{\infty}d\!\mu(r)
\frac{1}{2\pi i}\oint_{C_{z_i,r}}F
\,d\!z .
\end{equation}
More explicitly, this residue can be given as
\begin{equation}
{\rm Res}_{z= z_0}\frac{g(z)}{|z-z_0|^{2\alpha}}=
\sum_{i=0}^{\infty}\frac{\Lambda(i-\alpha)}{i!(i+1)!}
\partial_{z}^k\partial_{\bar{z}}^{k+1}g_{|_{z=z_0}}
\hspace{.5em}
(\alpha\in{\bf R}) .
\label{res}\end{equation}
Here $g$ is a  function defined and regular in some environment
of $z_0$. Note that the sum on the right-hand side of~(\ref{res})
always
has a
finite number of nontrivial  terms due to the property~(\ref{bound})
of
$\Lambda$.
Using the fact that  a contour integral of a full differential is
trivial,
it can be shown that
\begin{equation}\label{fd}
{\rm Res}_{z=z_0}\partial_{z} F={\rm
Res}_{\bar{z}=\bar{z}_0}\partial_{\bar{z}} F.
\end{equation}
Applying consequently~(\ref{rt}),  we can rewrite the first term on
the right-hand side of~(\ref{sc}) as
\begin{eqnarray}
{\cal R}_{z_n=\cdots=z_0}
\frac{1}{\pi}\int_{\Sigma}F\,d^2\!z
&=&
{\cal R}_{z_n=\cdots=z_0}
\left(
\frac{1}{2\pi i}\oint_{\partial\Sigma}
\partial_{\bar{z}}^{-1}F\,d\!z
- \sum_{i=0}^N{\rm Res}_{z=z_i}\partial_{\bar{z}}^{-1}F
\right)
\nonumber\\[1ex]
&=&
\frac{1}{2\pi i}\oint_{\partial\Sigma}
{\cal R}_{z_n=\cdots=z_0}{\partial_{\bar{z}}^{-1}F}
\,d\!z
\nonumber\\[1ex] &&
-\sum_{i=0}^N{\cal R}_{z_n=\cdots=z_0}
{\rm Res}_{z=z_i}\partial_{\bar{z}}^{-1}F
\nonumber\\[1ex]
&=&
\frac{1}{\pi}\int_{\Sigma}
{\cal R}_{z_n=\cdots=z_0}F
\,d^2\!z
+{\rm Res}_{z=z_0}{\cal R}_{z_n=\cdots=z_0}
\partial_{\bar{z}}^{-1}F
\nonumber\\[1ex]
&&
-\sum_{i=1}^N{\cal R}_{z_n=\cdots=z_0}
{\rm Res}_{z=z_i}\partial_{\bar{z}}^{-1}F
\nonumber\end{eqnarray}
and use for successors the formula
\begin{equation}
{\cal R}_{z\doteq z_n=\cdots=z_0}F=
{\rm Res}_{z=z_0}{\cal R}_{z_n=\cdots=z_0}
\partial_{\bar{z}}^{-1}F
-\sum_{i=0}^N{\cal R}_{z_n=\cdots=z_0}{\rm Res}_{z=z_i}
\partial_{\bar{z}}^{-1}F.
\label{sc1}\end{equation}
Here $\partial_{\bar{z}}^{-1}F$ is a function on
$\Sigma^{n+2}\setminus\Delta(\Sigma^{n+2})$
such as $\partial_{\bar{z}}\partial_{\bar{z}}^{-1}F=F$.
In fact, a function $\partial_{\bar{z}}^{-1}F$ defined in some
open environment of the main diagonal $z=z_n=\cdots z_0$
can be used in~(\ref{sc1}) as well.
We can add to $\partial_{\bar{z}}^{-1}F$ any function meromorphic
with respect to
$z$, and it will not affect the right-hand side of this formula.
Let us apply~(\ref{sc1}) to calculate a successor of the
antiholomorphic
derivative. Then taking into account
translation invariance of the generalized residue we will have
\begin{eqnarray}
\partial_{\bar{z}\doteq\bar{z}_0}F
&=&
{\rm Res}_{z=z_0}\partial_{z_0}\partial_{z}^{-1}F
-\partial_{z_0}{\rm Res}_{z=z_0}
\partial_{\bar{z}}^{-1}F
\nonumber\\[1ex]
&=&
{\rm Res}_{z=z_0}\partial_{z_0}
\partial_{z}^{-1}F-{\rm Res}_{z= z_0}
\left(
\partial_{z_0}+\partial_{\bar{z}}
\right)
\partial_{\bar{z}}^{-1}F
\nonumber\\[1ex]&=&
-{\rm Res}_{z=z_0}F.
\label{scd}\end{eqnarray}
Therefore, the antiholomorphic derivative is deformed as
\begin{equation}\label{dhd}
\delta\partial_{\bar{z}}\Upsilon=-{\rm
Res}_{z^{\prime}=z}\Psi(z^{\prime})\Upsilon(z).
\end{equation}
In the case of the cutoff regularization,
this result can be easily
understood as an effect of the deformation of the
integration area.
A more detailed technique of the successor calculation
 can be found in Appendix~\ref{appx}.
\sect{Finite Deformations}
Let us use,  for deformed amplitudes the following formula
\begin{equation}\label{texp}
{\langle 0 \rangle}_{\Sigma}^{\Psi} =
\left\langle
\exp\frac{1}{\pi}\int_{\Sigma}\Psi
\,d^2\!z
\right\rangle_{\Sigma}.
\end{equation}
Here $\Psi$ is some vertex operator function (not necessarily
primary),
and the contact divergences are regularized by the
method~(\ref{reg}).
Then the sewing property (axiom~\ref{sew}) will be automatically
satisfied, and only the condition of conformal invariance
will remain to be implemented.
Here and afterward we mark all the deformed objects
with the superscript symbol of the vertex operator function
parametrizing the deformation.
We will identify vertex operators of the deformed and initial
theories, by means of the formula
\begin{equation}\label{dv}
\langle \Upsilon(z_0)\rangle_{\Sigma}^{\Psi}=
\left\langle
\Upsilon(z_0)\exp
\frac{1}{\pi}\int_{\Sigma}\Psi(z)
\,d^2\!z
\right\rangle_{\Sigma} .
\end{equation}
Then the  T-product~(\ref{t-p}) for the
deformed theory will be
\begin{equation}
\langle
\Upsilon_0(z_0)\cdots\Upsilon_n(z_n)
\rangle_{\Sigma}^{\Psi}
=\left\langle
\Upsilon_0(z_0)\cdots\Upsilon_n(z_n)
\exp\frac{1}{\pi}\int_{\Sigma}\Psi(z)
\,d^2\!z
\right\rangle_{\Sigma} .
\label{dtp}\end{equation}
The formulas~(\ref{texp}),~(\ref{dv}) and~(\ref{dtp})
generalize the
analogous
formulas~(\ref{infdef}),~(\ref{infvrt})
and~(\ref{inftp}) for infinitesimal
deformations.

Let us consider a family of deformed theories associated with
a scaled  vertex operator function $\tau\Psi$ $(\tau\in{\bf R})$.
It is easy to see that the derivative of the corresponding
$T$-product~(\ref{dtp}) with respect to $\tau$
can be given as
\begin{equation}\label{de}
\frac{d}{d\tau}{\langle 0 \rangle}_{\Sigma}^{\tau\Psi}=
\frac{1}{\pi}\int_{\Sigma}
\langle
\Psi(z)
\rangle_{\Sigma}
^{\tau\Psi}
\,d^2\!z .
\end{equation}
Therefore, we can use the formulas~(\ref{rd}) to calculate
derivatives
and higher derivatives of the deformed residuelike operations
with respect to $\tau$.
Substituting them to the Taylor expansion
$$
{\cal R}^{\Psi}=\sum_{i=0}^{\infty}\frac{1}{i!}
\left(
\frac{\partial}{\partial\tau}
\right)^{i}
{\cal R}^{\tau\Psi}
\nolinebreak
\begin{picture}(33,7)(0,0)
\setlength{\unitlength}{.08em}\put(1,-7){\line(0,1){14}}
\put(2,-7){$\scriptstyle\tau=0$}
\end{picture}
,
 $$
we will come to the following perturbative formula for the finite
deformation:
\begin{eqnarray}
{\cal R}_{z_n=\cdots=z_0}^{\Psi}\Upsilon_0(z_0)
\cdots\Upsilon_n(z_n)
&=&\sum_{i=0}^{\infty}\frac{1}{i!}
{\cal R}_{z_{n+i}\doteq\cdots\doteq\bar{z}_1=\bar{z}}
\Upsilon_0(z_0)\cdots\Upsilon_n(z_n)
\nonumber\\&&\hspace{2em}\times
\Psi(z_{n+1})\cdots\Psi(z_{n+i}).
\label{fdr}\end{eqnarray}
In particular, for the deformed holomorphic and antiholomorphic
differentials we will have
\begin{eqnarray}
\partial_{z}^{\Psi}\Upsilon&=&
\partial_{z}\Upsilon
-\sum_{i=1}^{\infty}\frac{1}{i!}
{\rm Res}_{\bar{z}_i\doteq\cdots\doteq\bar{z}_1=\bar{z}}
\Upsilon(z)
\Psi(z_1)\cdots\Psi(z_i),
\nonumber\\[1ex]
\partial_{\bar{z}}^{\Psi}\Upsilon&=&
\partial_{\bar{z}}\Upsilon
-\sum_{i=1}^{\infty}\frac{1}{i!}
{\rm Res}_{z_i\doteq\cdots\doteq z_1=z}
\Upsilon(z)
\Psi(z_1)\cdots\Psi(z_i).
\label{fdd}\end{eqnarray}
Note that their commutator is kept to be trivial
$$
\left[
\partial_{z}^{\Psi},\partial_{\bar{z}}^{\Psi}
\right]
=0.
 $$
Therefore, this also can be used as a method to construct the Lax
pair.

After we regularized the contact divergences in the
$T$-exponent~(\ref{dtp}),
the boundary divergence still remained and even became stronger in
the higher orders of the approximation.
Of course, we can regularize the boundary divergence analogously as
we
did the contact divergences, but then it is  difficult to implement
the condition of conformal invariance and,
what is more important, it becomes impossible to satisfy it.
Conformally invariant infinitesimal deformations corresponding to
primary
fields, with the only exceptions of some specific cases, cannot be
corrected to restore conformal invariance even in the second
approximation.
However, the condition of conformal invariance with
respect to the fixed functor is, in fact, too strong, as the functor
itself may  be deformed.
If the boundary regularization
is not specified, the field $\Psi$ defines amplitudes
only modulo the group of local multipliers~(\ref{lm}).
Therefore, the condition of conformal
invariance can be applied only modulo this group,
{\em i.e\/}, modulo boundary term.

Let us fix regularization for
one of the boundary components of some
specific
Riemann surfaces. Then we can resolve the ambiguity of amplitudes
for all the rest of Riemann surfaces requiring the sewing property
(axiom ~\ref{sew}) to be satisfied.
The ambiguity will still remain if the deformed theory has
symmetries.
Otherwise, all the amplitudes will be projectively defined.
Then, we can define the deformed functor, requiring that
the operators~(\ref{fr}) representing maps between contours
are deformed by means of a product
with local multipliers~(\ref{lm}),
thus making the deformed amplitudes conformally invariant.
The energy-momentum tensor for such a deformed functor
can be shown to be
\begin{equation}\label{em}
T_{zz}^{\Phi}=T_{zz}+\Phi_{zz},\hspace{1em}
T_{\bar{z}\bar{z}}^{\Phi}=T_{\bar{z}\bar{z}}+\Phi_{\bar{z}\bar{z}}.
\end{equation}
Here $\Phi_{zz}$, $\Phi_{\bar{z}\bar{z}}$ are some normalizable
vertex
operator
functions.
Note, that the components of the energy-momentum tensor themselves
are
not normalizable.
The parameters $\Phi_{zz}$, $\Phi_{\bar{z}\bar{z}}$ depend on the
deformation
of the functor
and the method of regularization.

Applying~(\ref{hol}) to the deformed energy-momentum
tensor~(\ref{em})
\begin{equation}\label{eq}
\partial_{\bar{z}}^{\Psi} T_{zz}^{\Phi}=\partial_{z}^{\Psi}
T_{\bar{z}\bar{z}}^{\Phi}=0
\end{equation}
and using the formula~(\ref{fdd})
for the deformed derivatives,
we will come
to an equation on $\Psi$, $\Phi_{zz}$, $\Phi_{\bar{z}\bar{z}}$.
These equations and vertex operator functions
$\Psi$, $\Phi_{zz}$, $\Phi_{\bar{z}\bar{z}}$ can be
interpreted as equations of motion and dynamic fields.
The existence of  $\Phi_{zz}$ and $\Phi_{\bar{z}\bar{z}}$ satisfying
this
equation is,
in fact, a criteria  for the deformed theory to be conformally
symmetrical.

We can calculate the deformed representation of the Virasoro algebra
substituting the deformed energy-momentum tensor~(\ref{em})
and the deformed residue~(\ref{fdr}) to  the formula~(\ref{rep}):
\begin{eqnarray}
L_{v^z}^{\Psi}\Upsilon(z_0)&=&
{\rm Res}_{z=z_0}^{\Psi}v^z(z)T_{zz}^{\Phi}(z)\Upsilon(z_0)
\nonumber\\
&=&\sum_{i=0}^{\infty}\frac{1}{i!}
{\rm Res}_{z_i\doteq\cdots\doteq z_1\doteq z=z_0}
v^z(z) T_{zz}^{\Phi}(z)\Upsilon(z_0)
\Psi(z_i)\cdots\Psi(z_1) ,
\nonumber\\[1ex]
\bar{L}_{v^{\bar{z}}}^{\Psi}\Upsilon(z_0)&=&
{\rm Res}_{\bar{z}=\bar{z}_0}^{\Psi}
v^{\bar{z}}(z)
T_{\bar{z}\bar{z}}^{\Phi}(z)\Upsilon(z_0)
\nonumber\\
&=&\sum_{i=0}^{\infty}\frac{1}{i!}
{\rm Res}_{\bar{z}_i\doteq\cdots\doteq\bar{z}_1\doteq\bar{z}
=\bar{z}_0}
v^{\bar z}(z)T_{\bar{z}\bar{z}}^{\Phi}(z)\Upsilon(z_0)
\Psi(z_i)\cdots\Psi(z_1).
\nonumber\\
\label{dva}\end{eqnarray}
The commutative relations for the deformed Virasoro operators
\begin{eqnarray}&&
[L_n^{\Psi},L_k^{\Psi}]=(k-n)L_{k+n}^{\Psi} +
\frac{D}{12}\delta_{k+n,0}n(n^2-1),
\nonumber\\&&
[\bar{L}_{n}^{\Psi},\bar{L}_{k}^{\Psi}]=
(k-n)\bar{L}_{k+n}^{\Psi} +
\frac{D}{12}\delta_{k+n,0}n(n^2-1)\nonumber\\&&
[L_n^{\Psi},\bar{L}_{k}^{\Psi}]=0
\nonumber\end{eqnarray}
are equivalent to
\begin{eqnarray}
0&=&{\rm Res}_{z^{\prime}=z}^{\Psi}T_{zz}^{\Phi}(z^{\prime})
T_{zz}^{\Phi}(z)(z^{\prime}-z)^{k}+
\delta_{k,0}\partial_{z} T_{zz}^{\Phi}
+\delta_{k,1}T_{zz}^{\Phi}
-\frac{D}{2}\delta_{k,3}
\nonumber\\
&=&
{\rm Res}_{\bar{z}^{\prime}=\bar{z}}^{\Psi}
T_{\bar{z}\bar{z}}^{\Phi}(\bar{z}^{\prime})
T_{\bar{z}\bar{z}}^{\Phi}(z)(\bar{z}^{\prime}-\bar{z})^{k}+
\delta_{k,0}\partial_{\bar{z}}
T_{\bar{z}\bar{z}}^{\Phi}
+\delta_{k,1}T_{\bar{z}\bar{z}}^{\Phi}
-\frac{D}{2}\delta_{k,3}\nonumber\\&=&
{\rm Res}_{z^{\prime}=z}^{\Psi}T_{\bar{z}\bar{z}}^{\Phi}(z^{\prime})
T_{zz}^{\Phi}(z)(z^{\prime}-z)^{k}
={\rm Res}_{\bar{z}^{\prime}=\bar{z}}^{\Psi}
T_{\bar{z}\bar{z}}^{\Phi}(z^{\prime})
T_{zz}^{\Phi}(z)(z^{\prime}-z)^{k}
\ (k\geq 0) .
\nonumber\end{eqnarray}
Let us show that these equations are satisfied.
Indeed, the vertex operator functions in their left
parts are (anti)holomorphic, and,
as a consequence of~(\ref{em}), normalizable.
Nontrivial normalizable (anti)holomorphic modes
correspond to
symmetries of the deformed theory, which
form a Kac-Moody algebra, as in the Wess-Zumino-Witten model.
Presence of such symmetries is a specific, in some sense,
degenerate case.
In the usual situation
holomorphic modes do not exist and, therefore, the equations are
automatically satisfied. For symmetric phase we can satisfy them
adding to $T_{zz}^{\Phi}$ some holomorphic vertex operator function.

\subsection{Symmetries}
The vertex operator functions $\Psi$, $\Psi^{\prime}$ parametrize
equivalent
theories if corresponding amplitudes~(\ref{texp})
are similar modulo the group of local
multipliers~(\ref{lm}),
or, what is the same, modulo the boundary term
$$
{\langle 0 \rangle}_{\Sigma}^{\Psi^{\prime}}\sim{\langle 0
\rangle}_{\Sigma}^{\Psi}.
 $$
The similarity transformations of amplitudes induce
a  {\em covariant\/} transformation
of vertex operators
$$
\Upsilon\ \longrightarrow \ \Upsilon^{\prime}:\hspace{1em}
\langle
\Upsilon^{\prime}
\rangle_{\Sigma}
^{\Psi^{\prime}}\sim
\langle \Upsilon
\rangle_{\Sigma}^{\Psi} .
 $$
Infinitesimal symmetry transformations of the initial theory can be
described as
\begin{eqnarray}&&
\Psi\ \longrightarrow \ \Psi+\delta_{\xi}\Psi,\hspace{1em}
\Upsilon\ \longrightarrow \
\Upsilon+\hat{\xi}\Upsilon:
\nonumber\\[1ex]&&
\hat{\xi}\Upsilon(z_0)=
{\rm Res}_{\bar{z}=\bar{z}_0}\xi_z(z)\Upsilon(z_0)+
{\rm Res}_{z=z_0}\xi_{\bar{z}}\Upsilon(z_0),
\nonumber\\[1ex] &&
\delta_{\xi}\Psi=\partial_{\bar{z}}\xi_z+\partial_{z}\xi_{\bar{z}}
\hspace{1em}(\Psi=0).
\label{0sym}\end{eqnarray}
Here $\xi=(\xi_z,\xi_{\bar{z}})$ is a pair of vertex operator
functions,
parametrizing the symmetries.
The increment of the $T$-exponent~(\ref{dtp})
under the transformation~(\ref{0sym}) in a linear approximation
can be given as
\begin{eqnarray}
&&\hspace{-2em}\delta_{\xi}
\left(
\Upsilon(z_0)\exp\frac{1}{\pi}\int\Psi(z)\,d^2\!z
\right)
\approx
\frac{1}{\pi}\int\!d^2\!z_1\Psi(z_1)
{\rm Res}_{z_2=z_0}\xi_z(z_2)\Upsilon(z_0)
\nonumber\\&&+\frac{1}{2\pi^2}\int\!d^2\!z_1\int\!d^2\!z_2
\Upsilon(z_0)
\Bigl(
\Psi(z_1)\partial_{\bar{z}}\xi_z(z_2)
+\partial_{\bar{z}}\xi_z(z_1)\Psi(z_2)
\Bigr) +\, z\leftrightarrow\bar{z}
\nonumber\\[1ex]&=&
\frac{1}{\pi}\int\!d^2\!z_1
\Biggl(
\frac{1}{2}\Psi(z_1){\rm Res}_{z_2=z_0}
\xi_z(z_2)\Upsilon(z_0)
-\Upsilon(z_0){\rm Res}_{z_2=z_1}\xi_z(z_2)\Psi(z_1)
\Biggr.\nonumber\\
&&\Biggl.
-\partial_{z_1}\frac{1}{\pi}\int
d^2\!z_2\,\Upsilon(z_0)\xi_z(z_1)\Psi(z_2)+
\Upsilon(z_0){\rm Res}_{z_2=z_1}\Psi(z_2)\xi_z(z_1)
\Biggr)
+\, z\leftrightarrow\bar{z}
\nonumber\\[1ex]
&=&\frac{1}{\pi}\int\!d^2\!z_1
\left(
\frac{1}{2}\Psi(z_1){\rm Res}_{z_2=z_0}\xi_z(z_2)\Upsilon(z_0)
-\Upsilon(z_0){\rm Res}_{z_2=z_1}\xi_z(
\hspace{0.5em}\underbrace{\hspace{-0.5em}
z_2)\Psi(z_1
\hspace{-0.5em}}_{asym} \hspace{0.5em})
\right)
\nonumber\\[1ex]&&
-\frac{1}{2}{\rm Res}_{z_1=z_0}
\int\Upsilon(z_0)\xi_z(z_1)\Psi(z_2)\,d^2\!z_2
+\, z\leftrightarrow\bar{z}
\nonumber\\[1ex]&=&
-\int\Upsilon(z_0){\rm Res}_{z_2=z_1}\xi_z(
\hspace{0.5em}\underbrace{\hspace{-0.5em}
z_2)\Psi(z_1
\hspace{-0.5em}}_{asym} \hspace{0.5em}
)\,d^2\!z_1
-\frac{1}{2}{\rm Res}_{z_2\doteq
z_1=z_0}\Psi(z_2)\xi(z_1)\Upsilon(z_0)
\nonumber\\&&
 +\, z\leftrightarrow\bar{z} .
\nonumber\end{eqnarray}
Here we used formulas~(\ref{rt}) and~(\ref{sc})
and disregarded the boundary term.
For brevity
we omitted the angular brackets
symbolizing the  $T$-product.
This increment is trivial if $\Psi=0$;  {\em i.e.\/},     for
deformations of
the initial theory. If $\Psi\neq 0$,  it can be trivialized by the
following first order correction to the symmetry
transformation~(\ref{0sym}):
\begin{eqnarray}
\delta_{\xi}\Psi(z)&=&\partial_{\bar{z}}\xi_z(z)+{\rm
Res}_{z_1=z}\xi_z(
\hspace{0.5em}\underbrace{\hspace{-0.5em}
z_1)\Psi(z
\hspace{-0.5em}}_{asym} \hspace{0.5em}
)+ z\leftrightarrow\bar{z} + O(\Psi^2)\nonumber\\
\hat{\xi}\Upsilon(z_0)
&=&{\rm Res}_{\bar{z}=\bar{z}_0}\xi_z(z)\Upsilon(z_0)+
\frac{1}{2}{\rm Res}_{z_2\doteq
z_1=z_0}\Psi(z_2)\xi(z_1)
\Upsilon(z_0)
\nonumber\\ &&
+z\leftrightarrow\bar{z} + O(\Psi^2)  .
\label{sym}\end{eqnarray}
The higher order corrections to this transformation can be
calculated analogously. However, we cannot write an explicit formula
or give
an explicit procedure for their calculations yet.

Note that the  symmetries described above do not correspond
directly to the similarity transformations, which also depend on
$\Psi$ and
the boundary regularization.
Therefore, the commutator of such symmetries may be field dependent.
In other words, the symmetry algebra should not be closed.
In some sense, it is similar to a gauge-fixed Yang-Mills theory.

It might be asked what relationship the symmetries elucidated here
have
to the  suggestion of Banks and Martinec~\cite{BM} that renorm group
redundancies
may contribute to the symmetry algebra.
In our formalism,  such redundancies
can be interpreted  as a simultaneous changing of regularization
parameters
$\Lambda(\alpha)$ together with the vertex operator function $\Psi$
in
such a way
that it does not affect an equivalence class of deformed theory.
In fact, using such changing regularizations symmetries may help to
close
the symmetry algebra. Then we could combine vertex operator function
$\Psi$
and regularization parameters to unique covariant object
parametrizing two-dimensional field theories.
However,  our attempts to do it resulted in an excessive number of
auxiliary fields, that made the symmetry algebra as wide as it would
be if we simply extended it with field dependent transformations.

We don't think that it is a physical problem.
It just means that a classification of physical states as elements of
a quotient space for some closed symmetry group does  not work in
the string theory.

\subsection{Translationally invariant theories}
The regularization of contact divergences~(\ref{reg}) is not
conformally
invariant,
which complicates the implementation of the conformally invariance
for deformed amplitudes.
However, this regularization is still translationally invariant.
Therefore, if $\Psi$ is a translationally invariant  vertex operator
function;  {\em i.e.\/},
obeys~(\ref{transl}),
it will correspond to a translationally symmetrical theory.
This let us reduce the space of dynamic fields to
the space of translationally invariant vertex operator functions.

Then the deformed translation operators
$L_{-1}^{\Psi}$, $\bar{L}_{-1}^{\Psi}$ will be always defined,
whether $\Psi$ obeys the equation of motion~(\ref{eq}) or not.
They can be given as
\begin{eqnarray}
L_{-1}^{\Psi}\Upsilon(z)&=&{\rm
Res}_{z^{\prime}=z}T_{zz}^{\Psi}(z^{\prime})\Upsilon(z)+
{\rm Res}_{\bar{z}^{\prime}=z}
{T_{z\bar{z}}^{\Psi}}(z_1)\Upsilon(z)
\nonumber\\[1ex]
\bar{L}_{-1}^{\Psi}\Upsilon(z)&=&{\rm Res}_{\bar{z}^{\prime}=\bar{z}}
{T_{\bar{z}\bar{z}}^{\Psi}}(z^{\prime})\Upsilon(z)+
{\rm Res}_{z^{\prime}=z}T_{\bar{z} z}^{\Psi}(z^{\prime})\Upsilon(z).
\label{to}\end{eqnarray}
Here $T^{\Psi}$ is a nonconformal energy-momentum tensor, satisfying
\begin{eqnarray}&&\hspace{-1em}
\partial_{\bar{z}}^{\Psi}T_{zz}^{\Psi}+
\partial_{z}^{\Psi}T_{z\bar{z}}^{\Psi}=
\partial_{z}^{\Psi}T_{\bar{z}\bar{z}}^{\Psi}+
\partial_{\bar{z}}^{\Psi}T_{\bar{z}\bar{z}}^{\Psi}=0,
\nonumber\\[1ex]&&\hspace{-1em}
T_{zz}^{\Psi}\cong T_{zz},\hspace{1em}
T_{\bar{z}\bar{z}}^{\Psi}\cong T_{\bar{z}\bar{z}},\hspace{1em}
T_{z\bar{z}}^{\Psi}\cong T_{\bar{z}\bar{z}}^{\Psi}\cong 0.
\label{cl}\end{eqnarray}
Here the symbol $\cong$ indicates that the difference between
expressions at its left and right is a normalizable
vertex operator function.

A vertex operator function translationally invariant for initial
theory
remains translationally invariant for deformed theory.
Therefore,
\begin{equation}\label{ti}
\partial_{z}^{\Psi}+L_{-1}^{\Psi}=
\partial_{z}+L_{-1},\hspace{1em}
\partial_{\bar{z}}^{\Psi}+\bar{L}_{-1}^{\Psi}=
\partial_{\bar{z}}+\bar{L}_{-1} .
\end{equation}
Using here~(\ref{fdd}) we will come to the following  expression for
the deformed translation operators:
\begin{eqnarray}
L_{-1}^{\Psi}\Upsilon &=&L_{-1}\Upsilon
+\sum_{i=1}^{\infty}\frac{1}{i!}
{\rm Res}_{\bar{z}_i\doteq\cdots\doteq\bar{z}_1=\bar{z}}
\Psi(z_1)\cdots\Psi(z_i)\Upsilon(z),
\nonumber\\
\bar{L}_{-1}^{\Psi}\Upsilon&=&\bar{L}_{-1}\Upsilon
+\sum_{i=1}^{\infty}\frac{1}{i!}
{\rm Res}_{z_i\doteq\cdots\doteq
z_1=z}\Psi(z_1)\cdots\Psi(z_i)\Upsilon(z).
\label{fdto}\end{eqnarray}
Let us substitute $\Psi$ in the first equation of~(\ref{cl}) by
$\tau\Psi$
and then differentiate it with respect to $\tau$:
\begin{eqnarray}
0&=&-{\rm Res}_{z_1=z}^{\tau\Psi}\Psi(z_1)T_{zz}^{\Psi}(z)
-{\rm Res}_{\bar{z}_1=
\bar{z}}^{\tau\Psi}{\Psi}(z_1)T_{z\bar{z}}^{\Psi}(z)
\nonumber\\[1ex] &&
+\partial_{\bar{z}}^{\tau\Psi}
\frac{\partial}{\partial\tau}T_{zz}^{\tau\Psi}(z)+
\partial_{z}^{\tau\Psi}\frac{\partial}{\partial\tau}
T_{z\bar{z}}^{\tau\Psi}(z) .
\end{eqnarray}
Using  translation invariance of $\Psi$ here in the form
$$
0=\partial_{z}^{\tau\Psi}\Psi(z)+{\rm Res}_{z_1=z}
T_{zz}^{\tau\Psi}(z_1)\Psi(z)
+{\rm Res}_{z_1=z}T_{z\bar{z}^{\Psi}}(z_1)\Psi(z),
 $$
we will come to the equation:
\begin{eqnarray}
0&=&
\partial_{\bar{z}}^{\tau\Psi}
\frac{\partial}{\partial\tau}T_{zz}^{\tau\Psi}(z)+
\partial_{z}^{\tau\Psi}
\frac{\partial}{\partial\tau}
T_{z\bar{z}}^{\tau\Psi}(z)
-\partial_{z}^{\tau\Psi}
\Psi(z)
\nonumber\\&&
-2{\rm Res}_{z_1=z}^{\tau\Psi}\Psi(
\hspace{0.5em}\underbrace{\hspace{-0.5em}
z_1)T_{zz}^{\tau\Psi}(z
\hspace{-0.5em}}_{sym} \hspace{0.5em}
)-2{\rm Res}_{\bar{z}_1=\bar{z}}^{\tau\Psi}{\Psi}(
\hspace{0.5em}\underbrace{\hspace{-0.5em}
z_1)T_{z\bar{z}}^{\tau\Psi}(z
\hspace{-0.5em}}_{sym} \hspace{0.5em}
) .
\nonumber\end{eqnarray}
This equation will be satisfied if
\begin{eqnarray}
\frac{\partial}{\partial\tau}T_{zz}^{\tau\Psi}(z)
&=&
{\cal B}_{z_1=z}^{\tau\Psi}\Psi(z_1)T_{zz}^{\tau\Psi}(z)+
{\cal A}_{\bar{z}_1=\bar{z}}^{\tau\Psi}
\Psi(z_1)T_{z\bar{z}}^{\tau\Psi}(z),\nonumber\\[1ex]
\frac{\partial}{\partial\tau}T_{z\bar{z}}^{\tau\Psi}(z)
&=&\Psi(z)
+{\cal A}_{z_1=z}^{\tau\Psi}\Psi(z_1)T_{zz}^{\tau\Psi}(z)
+{\cal B}_{\bar{z}_1=\bar{z}}^{\tau\Psi}\Psi(z_1)
T_{z\bar{z}}^{\tau\Psi}(z) .
\label{cf}\end{eqnarray}
Here ${\cal A}_{z_1=z}$, ${\cal B}_{z_1=z}$ are residuelike
operations satisfying
\begin{equation}\label{fdab}
\partial_{z}{\cal A}_{z_1=z}+\partial_{\bar{z}}{\cal B}_{z_1=z}={\rm
Res}_{z_1=z}+{\rm Res}_{z=z_1}.
\end{equation}
Such residuelike operations exist and can be defined as
\begin{eqnarray}
&&{\cal A}_{z_1=z}G
=\left\{
\begin{array}{ll}
\sum_{i=0}^{\infty}\frac{2^{1-i-2k}}{i!(i+k)!}\Lambda(i-\alpha)
\partial_{\bar{z}}^i\partial_{z}^{i+k-1}F(z)
 & (k\not\in 2{\bf Z},\ k\geq 1)\\[1ex]
0, &{\rm otherwise,}
\end{array}\right. \nonumber\\[2ex]
&&{\cal B}_{z_1=z}G
= \left\{
\begin{array}{ll}
\sum_{i=0}^{\infty}\frac{2^{1+2k-i-\alpha}}{i!(i-k)!}
\Lambda(i-k-\alpha)
\partial_{z}^i\partial_{\bar{z}}^{i-k-1}F &
(k\not\in 2{\bf Z},\ k\leq -1)\\[1ex]
0, &{\rm otherwise.}
\end{array}\right.
\nonumber\\[2ex]
&&\left(
G=F
\left(
\frac{z+z_1}{2}
\right)
|z_1-z|^{-2\alpha}(z_{-1}-z)^{-k-1}
\right)
\label{lh}\end{eqnarray}
Differentiating~(\ref{cf}) with respect to $\tau$
and using  initial conditions
$$
\tau=0:\hspace{1em}
T_{zz}^{\tau\Psi}= T_{zz},\hspace{1em}
T_{z\bar{z}}^{\tau\Psi}=0,
 $$
we can recurrently calculate all the higher derivatives
of the energy-momentum tensor
and then substitute them to  the Taylor expansion:
$$
T^{\Psi}=\sum_{i=0}^{\infty}\frac{1}{i!}
\left(
\frac{\partial}{\partial\tau}
\right)^i
T^{\tau\Psi}
\nolinebreak
\begin{picture}(33,7)(0,0)
\setlength{\unitlength}{.08em}\put(1,-7){\line(0,1){14}}
\put(2,-7){$\scriptstyle\tau=0$}
\end{picture} .
 $$
For example, in the second order approximation we have
\begin{eqnarray}
T_{zz}^{\Psi}(z)&=&T_{zz}(z)
+{\cal B}_{z_1=z}\Psi(z_1) T_{zz}(z)
+\frac{1}{2}{\cal B}_{z_2\doteq z_1=z}
\Psi(z_2)\Psi(z_1)T_{zz}(z)
\nonumber\\&&
+\frac{1}{2}{\cal B}_{z_2=z}{\cal B}_{z_1=z}\Psi(z_2)\Psi(z_1)
T_{zz}(z)
\nonumber\\[1ex]&&
+\frac{1}{2}{\cal A}_{\bar{z}_2=\bar{z}}
{\cal A}_{z_1=z}\Psi(z_2)\Psi(z_1)T_{zz}(z)
\nonumber\\ &&
+\frac{1}{2}{\cal A}_{\bar{z}_1=\bar{z}}\Psi(z_1)\Psi(z)
+0(\Psi^{3}) ,
\nonumber\\[2ex]
T_{z\bar{z}}^{\Psi}(z)&=&
\Psi(z)
+{\cal A}_{z_1=z}\Psi(z_1)T_{zz}(z)
+\frac{1}{2}{\cal A}_{z_2\doteq z_1=z}
\Psi(z_2)\Psi(z_1)T_{zz}(z)
\nonumber\\[1ex] &&+
\frac{1}{2}{\cal A}_{z_2=z}{\cal B}_{z_1=z}\Psi(z_2)\Psi(z_1)
T_{zz}(z)
\nonumber\\[1ex]&&
+\frac{1}{2}{\cal B}_{\bar{z}_2=\bar{z}}
{\cal A}_{z_1=z}\Psi(z_2)\Psi(z_1)T_{zz}(z)
\nonumber\\[1ex] &&
+\frac{1}{2}{\cal B}_{\bar{z}_1=\bar{z}}
\Psi(z_1)\Psi(z)+0(\Psi^{3}).
\label{ema}\end{eqnarray}
The analogous formulas for $T_{\bar{z}\bar{z}}^{\Psi}$,
$T_{\bar{z} z}^{\Psi}$ can be obtained by exchanging
the symbols $z$ and $\bar{z}$.

The symmetries~(\ref{sym}) remaining for translationally invariant
theories
can be parametrized by the translationally invariant vertex operator
functions $\xi_z$, $\xi_{\bar{z}}$.
The choice of the energy-momentum tensor is not unique, as the adding
to its
components full differentials of translationally invariant
vertex operator functions
\begin{eqnarray}
T^{\Psi}\ \longrightarrow \  T^{\Psi,\Phi}:\ &&
T_{zz}^{\Psi,\Phi}=T_{zz}^{\Psi}-
\partial_{z}^{\Psi}\Phi_z,\hspace{1em}
T_{z\bar{z}}^{\Psi,\Phi}=
T_{z\bar{z}}^{\Psi}+\partial_{\bar{z}}^{\Psi}\Phi_z,
\nonumber\\
&&T_{\bar{z}\bar{z}}^{\Psi,\Phi}=
T_{\bar{z}\bar{z}}^{\Psi}-
\partial_{\bar{z}}^{\Psi}\Phi_{\bar{z}},
\hspace{1em}
T_{\bar{z} z}^{\Psi,\Phi}=T_{\bar{z} z}^{\Psi}
+\partial_{z}^{\Psi}\Phi_{\bar{z}}
\label{emc}\end{eqnarray}
will not
affect the corresponding translation operators~(\ref{to}).
Therefore, its transformation under~(\ref{sym})
may differ from covariant  by such full differentials;  {\em i.e.\/},
$$
\delta_{\xi}T^{\Psi}\equiv
\frac{\delta\!T^{\Psi}}{\delta\Psi}\delta_{\xi}\Psi=
\hat{\xi}T^{\Psi}-H(\xi),
 $$
where
\begin{eqnarray}&&
H(\xi)_{zz}=-\partial_{z}^{\Psi}A(\xi)_z,\hspace{1em}
H(\xi)_{z\bar{z}}=\partial_{\bar{z}}^{\Psi}A(\xi)_z,
\nonumber\\[1ex]&&
H(\xi)_{\bar{z}\bar{z}}=-
\partial_{\bar{z}}^{\Psi}A(\xi)_{\bar{z}} ,
\hspace{1em}
H(\xi)_{\bar{z} z}=\partial_{z}^{\Psi}A(\xi)_{\bar{z}}.
\nonumber\end{eqnarray}
However, the modified energy-momentum tensor $T^{\Psi,\Phi}$
will transform covariantly,
if we define a transformation low for the fields $\Phi_z$,
$\Phi_{\bar{z}}$
by the formula
\begin{equation}\label{symd}
\Phi\ \longrightarrow \ \Phi+\hat{\xi}\Phi+A(\xi).
\end{equation}
As the vertex operator functions $\Phi_z$, $\Phi_{\bar{z}}$ are also
translationally invariant,  we can instead of~(\ref{emc}) use for
the covariant energy-momentum tensor an expression
\begin{eqnarray}
T_{zz}^{\Psi,\Phi}=T_{zz}^{\Psi}+L_{-1}^{\Psi}\Phi_z,\hspace{1em}&&
T_{z\bar{z}}^{\Psi,\Phi}=
T_{z\bar{z}}^{\Psi}-\bar{L}_{-1}^{\Psi}\Phi_z ,
\nonumber\\[1ex]
T_{\bar{z}\bar{z}}^{\Psi,\Phi}=
T_{\bar{z}\bar{z}}^{\Psi}+
\bar{L}_{-1}^{\Psi}\Phi_{\bar{z}},\hspace{1em}&&
T_{\bar{z} z}^{\Psi,\Phi}=T_{\bar{z}
z}^{\Psi}-L_{-1}^{\Psi}\Phi_{\bar{z}}  .
\label{emt}\end{eqnarray}
If the theory is conformally symmetrical, there exists $\Phi_z$,
$\Phi_{\bar{z}}$ trivializing the
contradiagonal components of $T^{\Psi,\Phi}$
\begin{equation}\label{eqg}
T_{z\bar{z}}^{\Psi,\Phi}=T_{\bar{z} z}^{\Psi,\Phi}=0.
\end{equation}
Then the diagonal components  of $T^{\Psi,\Phi}$
will form a conformal energy-\-mo\-men\-tum tensor.
They will be (anti)holo\-morphic
in a consequence of~(\ref{cl}).
Therefore, the equation~(\ref{eqg}) and the translationally
invariant vertex operator functions $\Psi$, $\Phi_z$ and
$\Phi_{\bar{z}}$
can be considered as an equation
of motion and dynamic fields for the closed string field theory.
This equation is equivalent to~(\ref{eq})
for theories without symmetries and may be a stronger requirement
otherwise.

\subsection{Linear approximation}
In the terms of~(\ref{ema}) corresponding to the first order
deformations
of
the energy-momentum tensor,
the residuelike operations~(\ref{lh}) are applied to the
functions holomorphic with respect to $z$.
By using for such functions the Loran expansion
$$
F=\sum_i(z^{\prime}-z)^{-i-1}{\rm
Res}_{z^{\prime}=z}(z^{\prime}-z)^iF,
 $$
it can be shown that
\begin{equation}
{\cal A}_{z^{\prime}=z}F =
\frac{1}{2}\sum_{i=1}^{\infty}
\frac{1}{k!}\partial_{z}^{i-1}
{\rm Res}_{z^{\prime}=z}(z^{\prime}-z)^i F,
\hspace{1em}
{\cal B}_{z^{\prime}=z}F=0.
\label{hlh}\end{equation}
Therefore, in the linear approximations components of
the deformed  energy-momentum tensor are equal to
\begin{equation}
T_{zz}^{\Psi}=T_{zz},\hspace{1em}
T_{z\bar{z}}^{\Psi}={\cal O}_{0}\Psi\hspace{1em}
T_{\bar{z}\bar{z}}^{\Psi}=T_{\bar{z}\bar{z}},\hspace{1em}
T_{\bar{z} z}^{\Psi}=\overline{\cal O}_{0}\Psi .
\label{eml}\end{equation}
Hereafter
\begin{equation}
{\cal O}_k=\delta_{k,0}+
\sum_{j=0}^{\infty}\frac{
(L_{-1})^j L_{k+j}}{(k+j+1)!},\hspace{1em}
\bar{{\cal O}_k}=\delta_{k,0}
+\sum_{j=0}^{\infty}\frac{(\bar{L}_{-1})^j
\bar{L}_{k+j}}{(k+j+1)!}.
\label{ooo}\end{equation}
States with bounded energy obey the condition
\begin{equation}\label{eb}
L_i\Psi=0\hspace{1em}(i\geq l),
\end{equation}
where $l$ is a level of string excitation.
For such states a number of nontrivial term in the sum in~(\ref{ooo})
is
always finite.

In the linear approximation we can substitute
the deformed translation operators $L_{-1}^{\Psi}$,
$\bar{L}_{-1}^{\Psi}$
in the formula~(\ref{emt})  for the covariant energy-momentum tensor
by the initial ones
\begin{eqnarray}
T_{z\bar{z}}^{\Psi,\Phi}={\cal O}_{0}\Psi
-\bar{L}_{-1}\Phi_z,\hspace{1em}&&
T_{\bar{z}z}^{\Psi,\Phi}
=\overline{\cal O}_{0}\Psi-L_{-1}\Phi_z,
\nonumber\\[1ex]
T_{zz}^{\Psi,\Phi}=T_{zz}+L_{-1}\Phi_z,\hspace{1em}&&
T_{\bar{z}\bar{z}}^{\Psi,\Phi}=
T_{\bar{z}\bar{z}}+\bar{L}_{-1}\Phi_{\bar{z}}.
\label{tl}\end{eqnarray}
Then, the linearized equation of motion~(\ref{eqg}) will be
\begin{equation}
{\cal O}_{0}\Psi=\bar{L}_{-1}\Phi_z,\hspace{1em}
\overline{\cal O}_{0}\Psi=L_{-1}\Phi_z.
\label{eql}\end{equation}
Let us calculate the covariant transformation of the energy-momentum
tensor
under linearized symmetries~(\ref{0sym})
\begin{eqnarray}
\hat{\xi}T_{zz}(z)&=&{\rm Res}_{z^{\prime}=z}
\xi_z(z^{\prime})T_{zz}(z)
+{\rm Res}_{\bar{z}^{\prime}=\bar{z}}
\xi_{\bar{z}}(z^{\prime})T_{zz}(z)
\nonumber\\[1ex]
&=&{\rm Res}_{z^{\prime}=z}
\sum_{i}(z-z^{\prime})^{i-1}
L_{1-i}\xi_z(z^{\prime})
\nonumber\\[1ex]&&
+{\rm Res}_{\bar{z}_1=\bar{z}}
\sum_{i}(z-z_1)^{i-1}L_{1-i}\xi_{\bar{z}}(z_1)
\nonumber\\
&=&L_{-1}{\cal O}_{0}\xi_z(z).
\nonumber\end{eqnarray}
Analogously,
$$
\hat{\xi}T_{\bar{z}\bar{z}}=\bar{L}_{-1}\overline{\cal
O}_{0}\xi_{\bar{z}}.
 $$
Substituting it together with~(\ref{eml}) in~(\ref{symd})
we will come to  the following formula
for the transformation of $\Phi$
under these symmetries:
\begin{equation}\label{sys0a}
\Phi_z\ \longrightarrow \ \Phi_z+{\cal O}_{0}\xi_z,\hspace{1em}
\Phi_{\bar{z}}\ \longrightarrow \ \Phi_{\bar{z}}+\overline{\cal
O}_{0}\xi_{\bar{z}}.
\end{equation}
Applying~(\ref{sch}) in~(\ref{dva}) and disregarding
the higher order terms,
we will come to the following formula for deformation
of the left Virasoro representation  in $H_{z_0}$:
\begin{eqnarray}
\delta L_{v^z}&=&
{\rm Res}_{z\doteq z_1=z_0}\Psi(z)v^zT_{zz}(z_1)
-{\rm Res}_{z=z_0}v^z\partial_{z}\Phi_z(z)\nonumber
\\[1ex]&=&
{\rm Res}_{\bar{z}=\bar{z}_0}J[v^z]_{\bar{z}}(z)
+{\rm Res}_{z=z_0}J[v^z]_z(z).
\label{dlv}\end{eqnarray}
Here
$$
J[v^z]_z=\Phi_z\partial_{z} v^z
 $$
and
\begin{eqnarray}
J[v^z]_{\bar{z}}&=&
\sum_{k=1}^{\infty}\frac{(-)^k}{k!}\partial_{z}^{k-1}
{\rm Res}_{z_1=z}(z_1-z)^k\Psi(z) v^z
T_{zz}(z_1)-v^z\partial_{\bar{z}}\Phi_z
\nonumber\\[1ex]
&=&
\sum_{k=1}^{\infty}
\sum_{i=0}^{\infty}\frac{(-)^k}{k!i!}\partial_{z}^{k-1}
{\rm Res}_{z_1=z}(z_1-z)^{k+i}
\partial_{z}^i v^z\Psi(z) T_{zz}(z_1)
-v^z\partial_{\bar{z}}\Phi_z
\nonumber\\[1ex]&=&
-\sum_{k=1}^{\infty}\sum_{i=0}^{\infty}
\frac{1}{i!k!}L_{-1}^{k-1}L_{k+i-1}\partial_{z}^i
v^z\Psi(z)-v^z\partial_{\bar{z}}\Phi_z\nonumber\\[1ex]
&=&
-\sum_{i=0}^{\infty}\frac{1}{i!}\partial_{z}^i v^z{\cal O}_{i}\Psi
+v^z(\Psi+\bar{L}_{-1}\Phi_z).
\nonumber\end{eqnarray}
Using the linearized equation of motion~(\ref{eql}),
we can also write it as
$$
J[v^z]_{\bar{z}}= v^z\Psi-\sum_{i=1}^{\infty}\frac{1}{i!}
{\cal O}_{i}\Psi\partial_z^i v^z.
 $$
Analogously, for the deformation of the right
Virasoro representation we have
\begin{equation}\label{drv}
\delta\bar{L}_{v^{\bar{z}}}=
{\rm Res}_{z=z_0}\bar{J}[v^{\bar{z}}]_z(z)+
{\rm Res}_{\bar{z}=\bar{z}_0}
\bar{J}[v^{\bar{z}}]_{\bar{z}}(z),
\end{equation}
where
$$
\bar{J}[v^{\bar{z}}]_z=
v^{\bar{z}}\Psi-\sum_{i=1}^{\infty}\frac{1}{i!}\partial_{\bar{z}}^i
v^{\bar{z}}\overline{\cal O}_{i}\Psi,\hspace{1em}
\bar{J}[v^{\bar{z}}]_{\bar{z}}=\partial_{\bar{z}} v^{\bar{z}}\Phi_z.
 $$
Because of~(\ref{eb}), the operators $J_z$, $J_{\bar{z}}$,
$\bar{J}_{\bar{z}}$, $\bar{J}_{z}$ act on tangent fields $v^{z}$,
$v^{\bar{z}}$
as differential operators of the order $l$. For the constant fields
$v_{-1}^z={v_{-1}}^{\bar{z}}\equiv 1$, they are
$$
J_{\bar{z}}=\bar{J}_z=\Psi,\hspace{1em} J_z=\bar{J}_{\bar{z}}=0
 $$
and, therefore,
$$
\delta\!L_{-1}=
{\rm Res}_{\bar{z}=\bar{z}_0}\Psi(z),\hspace{1em}
\delta\!\bar{L}_{-1} =
{\rm Res}_{z=z_0}\Psi(z).
 $$
Note that
the formula~(\ref{fdto}) taken in the linear approximations gives the
same
result.

According to the definition of nonholomorphic residue~(\ref{drlr}),
the deformation~(\ref{dlv}),~(\ref{drv}) of the Virasoro operators is
an
average of the
deformations
\begin{eqnarray}
\delta L_{v^z}&=&
\frac{1}{2\pi i}\oint_{\Gamma}J[v^z]_{\bar{z}}
\,d\!\bar{z}
+\frac{1}{2\pi i}\oint_{\Gamma}J[v^z]_z\,d\!z,
\nonumber\\[1ex]
\delta\bar{L}_{v^{\bar{z}}}
&=&
\frac{1}{2\pi i}\oint_{\Gamma}
\bar{J}[v^{\bar{z}}]_z\,d\!z +
\frac{1}{2\pi i}
\oint_{\Gamma}\bar{J}[v^{\bar{z}}]_{\bar{z}}
\,d\!\bar{z} ,
\label{uvd}\end{eqnarray}
taken over circular contours $\Gamma$.
It can be shown that solutions of~(\ref{eql}) satisfy
$$
L_{v^z}\Psi
=\partial_{z} J[v^z]_{\bar{z}}
+\partial_{\bar{z}}J[v^z]_z ,
\hspace{1em}
\bar{L}_{v^{\bar{z}}}\Psi=
\partial_{\bar{z}}\bar{J}[v^{\bar{z}}]_z
+\partial_{z}\bar{J}[v^{\bar{z}}]_{\bar{z}},
 $$
what can be interpreted
as conformal invariance of the deformed propagator~(\ref{infdef})
under the deformed representation of the Virasoro
algebra~(\ref{dlv}).
Such a simple interpretation of conformal invariance
cannot be applied beyond the linear approximation because of the
boundary divergence.

Equations~(\ref{eql}) are satisfied if $\Phi$ is trivial and
$\Psi$ obeys the conventional closed string
equation~(\ref{canon}). Then
$$
J[v^z]_{\bar{z}}=v^z\Psi
,\hspace{1em}\bar{J}[v^{\bar{z}}]_z=
v^{\bar{z}}\Psi,\hspace{1em}J[v^z]_z=
\bar{J}[v^{\bar{z}}]_{\bar{z}}=0,
 $$
and the deformation of the Virasoro representation~(\ref{uvd}) will
be
the same as~(\ref{mark}).
The relaxation of the equations of motion, what we have here,
is compensated by the symmetries~(\ref{0sym}) and~(\ref{sys0a})
and does not create additional physical degrees of freedom.
Analogous to~(\ref{dlv}) infinitesimal deformations of
the Virasoro representation corresponding to vertex operator
functions, which
are not primary fields, have been first found in~\cite{eg,eg1}
in the low-energy limit.

\sect{Spacetime Interpretation}
In this section
we will show how the definition of CFT given in
Section~\ref{axioms} deals
with the conventional path integral approach  and define the initial
(vacuum) CFT.
Then we will formulate a method for  nonperturbative analysis in
the low-energy limit and show how  it  corresponds to
Brans-Dicke theory of gravity interacting with a skew symmetric
tensor
field.

\subsection{Path integral approach}
Let us denote infinite dimensional manifolds of continuous
maps from the contour $\Gamma$ and the surface $\Sigma$ to the
$D$-dimensional manifold  ${\cal
M}$ (spacetime)  as ${\cal M}^{\Gamma}$ and ${\cal M}^{\Sigma}$,
respectively.
We can define ${\cal H}^{\Gamma}$ as a space of continuous functions
in ${\cal M}^{\Gamma}$,
endowed with the Hilbert product
$$
(\Psi,\Phi)=
\int_{{\cal M}^{\Gamma}}\bar{\Psi}
\left[
{\bf x}^{\Gamma}
\right]
\Phi
\left[
{\bf x}^{\Gamma}
\right]
d\, {\bf x}^{\Gamma} \ .
 $$
Then the conformally symmetrical amplitudes can be formally defined
through  the path integrals
\begin{equation}\label{pathe}
{\langle 0 \rangle}_{\Sigma}
\left[
{\bf x}_0^{\partial\Sigma}
\right]=
\int_{{\bf x}\nolinebreak\begin{picture}(10,3)(0,0)
\setlength{\unitlength}{.5pt}
\put(1,-7){\line(0,1){14}}
\put(2,-7){$\scriptscriptstyle\partial\Sigma$}
\end{picture}
={\bf x}_0}
\exp{
\left(
-S
\left[
{\bf x}^{\Sigma}
\right]
\right)
}
d\!{\bf x}^{\Sigma}\ .
\end{equation}
Here  $S$ is some conformally invariant action.
However, in general, this method to construct CFT's is not
quite correct.
First of all, the path integral procedure is rather ambiguous;
in addition, it can violate conformal invariance and be divergent.
An attempt to describe CFT in such a way leads to the well-known
$\beta$-function
approach, the shortcoming of which is mentioned in
the Introduction.
However, in the case of flat background;  {\em i.e.\/},
for the action
\begin{equation}\label{flat}
S=\frac{1}{\pi}\int_{\Sigma} G_{\nu\mu}
\partial x^{\nu}\bar{\partial} x^{\mu}\,d^2\! z
\end{equation}
with constant $G_{\nu\mu}$,
the path integral is Gaussian and can be explicitly calculated up to
a constant multiplier.
States having one-point support can be formally expressed through the
path integral
\begin{equation}\label{vp}
\Phi(z_0)[x^{\Gamma}]=
\int_{{\bf x}\nolinebreak
\begin{picture}(10,3)(0,0)
\setlength{\unitlength}{.5pt}
\put(1,-7){\line(0,1){14}}
\put(2,-7){$\scriptscriptstyle\partial\Sigma$}
\end{picture}
={\bf x}_0}
\phi(z_0)\exp{
\left(
-S\left[
{\bf x}^{\Sigma}
\right]
\right)
}
\, d\! {\bf x}^{\Sigma}.
\end{equation}
In the case of linear function $\phi=x^{\nu}$,
this integral can be reduced to Gaussian and also explicitly
calculated.
Corresponding vertex operators $X^{\nu}$ are operators of local
string  coordinates.
For the nonlinear $\phi$ integral~(\ref{vp}) is divergent.
We  will put into correspondence to such nonlinear functions normal
ordered
operators $:\!\phi\! :$, defined by means of the Wick formula with
the Green function
\begin{equation}\label{gf}
{\cal G}\left(
x^{\nu}(z),x^{\mu}(u)
\right)
=-2G^{\nu\mu}\ln|z-u|.
\end{equation}
Note that such normal ordering is not conformally invariant.

\subsection{Global deformations of spacetime metric}
Let us consider deformations corresponding to the vertex operator
function of the type
\begin{equation}\label{stgf}
\Psi=H_{\nu\mu}\partial X^{\nu}\bar{\partial} X^{\mu}(z^{\prime}).
 \end{equation}
Here  $H_{\nu\mu}$ is a matrix with constant coefficients.
CP-invariant deformed theories
(theories obeying Axiom~\ref{CP}) correspond  to Hermitian
matrices
$$
H_{\nu\mu}=\bar{H}_{\mu\nu}.
 $$
Such matrices can be given as
$$
H_{\nu\mu}=B_{\nu\mu}+iA_{\nu\mu},
 $$
where $B$ and $A$ are, respectively, symmetric and skew symmetric
real matrices.
Let us apply formula~(\ref{dhd})  to  calculate the deformation
of the coordinate differentials:
$$
\delta\partial X^{\eta}=-{\rm Res}_{\bar{z}^{\prime}=\bar{z}}
H_{\nu\mu}\partial X^{\nu}(z^{\prime})\bar{\partial}
X^{\mu}(z^{\prime})X^{\eta}(z) .
 $$
Here the T-product of the vertex operators under the residue has
the contact singularity
$$
\langle
\partial X^{\nu}(z^{\prime})\bar{\partial}
X^{\mu}(z^{\prime})X^{\eta}(z)
\rangle_{\Sigma}
\approx
-\frac{1}{z^{\prime}-z}G^{\nu\eta}
\langle
\bar{\partial} X^{\mu}(z^{\prime})
\rangle_{\Sigma}
-\frac{1}{\bar{z}^{\prime}-\bar{z}}G^{\mu\eta}
\langle
\partial X^{\nu}(z^{\prime})
\rangle_{\Sigma}
 $$
and, therefore,
\begin{equation}\label{stfi}
\delta\partial X^{\eta}=G^{\mu\eta}H_{\nu\mu}\partial X^{\nu} .
\end{equation}
The symbol "$\approx$" indicates that the difference between
expressions at its right and left is a regular function.

It can be shown that deformation of the $T$-product~(\ref{inftp})
corresponding to the
vertex operator function~(\ref{stgf}) does not affect contact
singularities of the coordinate
differentials. Therefore, deformation of such singularities is
determined by the deformations~(\ref{stfi})
\begin{eqnarray}
\delta
\langle
\partial X^{\alpha}(z_1)\partial X^{\beta}(z_2)
\rangle_{\Sigma}
&\approx&
\langle
\delta\partial X^{\alpha}(z_1)\partial X^{\beta}(z_2)
\rangle_{\Sigma}
+\langle
\partial X^{\alpha}(z_1)\delta\partial X^{\beta}(z_2)
\rangle_{\Sigma}
\nonumber\\[1ex]
&\approx& G^{\mu\alpha}H_{\nu\mu}
\langle
\partial X^{\nu}(z_1)\partial X^{\beta}(z_2)
\rangle_{\Sigma}
\nonumber\\[1ex]&&
+G^{\mu\beta}H_{\nu\mu}
\langle
\partial X^{\alpha}(z_1)\partial X^{\nu}(z_2)
\rangle_{\Sigma}
\nonumber\\[1ex]
&\approx& \frac{2}{(z_1-z_2)^2}G^{\mu\alpha}
G^{\nu\beta}B_{\nu\mu}{\langle 0
\rangle}_{\Sigma}.
\nonumber\end{eqnarray}
Analogously, for antiholomorphic coordinate differentials we have
\begin{eqnarray}
&&\delta\bar{\partial} X^{\eta}=G^{\nu\eta}H_{\nu\mu}\bar{\partial}
X^{\mu} ,\nonumber\\[1ex]
&&\delta\langle
\bar{\partial} X^{\alpha}(z_1)\bar{\partial} X^{\beta}(z_2)
\rangle_{\Sigma}=
\frac{2}{(z_1-z_2)^2}
G^{\mu\alpha}G^{\nu\alpha}B_{\nu\mu}{\langle 0 \rangle}_{\Sigma}.
\label{stfib}\end{eqnarray}
Thus, such deformed theory is still a theory in the flat background,
with a
modified metric
\begin{equation}\label{stfim}
g^{\alpha\beta}=G^{\alpha\beta}+\delta g^{\alpha\beta},\hspace{1em}
\delta g^{\alpha\beta}=2B_{\nu\mu}G^{\mu\alpha}G^{\nu\beta}.
\end{equation}
Therefore, the family of scaled fields $\tau H_{\nu\mu}\partial
X^{\nu}\bar{\partial}
X^{\mu}$
parametrize a family of  CFT's corresponding to different flat
metrics $g=g(\tau)$.
The deformations~(\ref{stfi}) and~(\ref{stfib}) of the coordinated
differentials are linear.
Therefore, the deformed coordinate differentials can be
expressed through the initial coordinate differentials,
$$
\partial^{\tau\Psi}X^{\eta}=U(\tau)_{\xi}^{\eta}\partial
X^{\xi},\hspace{1em}
\bar{\partial}^{\tau\Psi}X^{\eta}=
\left.
\tilde{U}(\tau)
\right._{\xi}^{\eta}\partial X^{\xi},
 $$
or, in the vector form,
\begin{equation}\label{stfe}
\partial^{\tau\Psi}{\bf X}=U(\tau)\partial{\bf X},\hspace{1em}
\bar{\partial}^{\tau\Psi}{\bf X}=\tilde{U}(\tau)\partial X^{\xi}.
\end{equation}
Here $U$,$\tilde{U}$ are some matrices.
Each of them can be used to calculate
the modified contravariant metric:
\begin{equation}\label{stffm}
g(\tau)=U(\tau)GU^{T}(\tau)=\tilde{U}(\tau)G\tilde{U}^{T}(\tau).
\end{equation}
Let us express the field $\Psi$ through deformed coordinate
differentials,
\begin{equation}\label{stfh}
\Psi=h(\tau)_{\nu\mu}
\partial^{\tau\Psi}X^{\nu}\bar{\partial}^{\tau\Psi}X^{\mu}
,\hspace{.5em}
h(\tau)=
(U^{-1})^{T}(\tau)H\tilde{U}^{-1}(\tau),
\end{equation}
and apply to the deform theory
formulas~(\ref{stfi}) and~(\ref{stfib}).
Then we will come to the differential equations
\begin{eqnarray}
\frac{\partial}{\partial\tau}
\partial^{\tau\Psi}{\bf X}
&=&g(\tau)h^{T}(\tau)\partial^{\tau\Psi}{\bf X}
=\tilde{U}GH^T\partial{\bf X},\nonumber\\[1ex]
\frac{\partial}{\partial\tau}
\bar{\partial}^{\tau\Psi}{\bf X}
&=&g(\tau)h(\tau)\bar{\partial}^{\tau\Psi}{\bf X}
=UGH\partial{\bf X},
\nonumber\end{eqnarray}
which can also  be written as
\begin{equation}\label{stfde}
\frac{\partial}{\partial\tau}U(\tau)=\tilde{U}(\tau)GH^T,\hspace{1em}
\frac{\partial}{\partial\tau}\tilde{U}(\tau)=U(\tau)GH.
\end{equation}
Solving these equations with the initial conditions
$$
U(\tau)\nolinebreak
\begin{picture}(33,7)(0,0)
\setlength{\unitlength}{.08em}
\put(1,-7){\line(0,1){14}}\put(2,-7){$\scriptstyle\tau=0$}
\end{picture}=
\tilde{U}(\tau)\nolinebreak
\begin{picture}(33,7)(0,0)
\setlength{\unitlength}{.08em}\put(1,-7){\line(0,1){14}}
\put(2,-7){$\scriptstyle\tau=0$}
\end{picture}=1
 $$
and putting then $\tau=1$ we can calculate $U$ and $\tilde{U}$,
\begin{eqnarray}
U&=&\cosh
\left(
\sqrt{GHGH^T}
\right)
+H^T\frac{\sinh
\left(
\sqrt{GHGH^T}
\right)
}
{\sqrt{GHGH^T}},
\nonumber\\[1ex]
\tilde{U}&=&\cosh(\sqrt{GH^TGH})+H\frac{\sinh
\left(
\sqrt{GH^TGH}
\right)
}
{\sqrt{GH^TGH}},
\nonumber\end{eqnarray}
and then substitute it to~(\ref{stffm}).
It gives the following formula for  the deformed metric:
\begin{eqnarray}
g&=&\frac{1}{2}\cosh
\left(
2\sqrt{HH^T}
\right)+
\frac{1}{2}\cosh
\left(
2\sqrt{H^TH}
\right)
\nonumber\\[1ex]&&+
H^T\frac{\sinh 2
\left(
\sqrt{HH^T}
\right)
}{2\sqrt{HH^T}}+
H\frac{\sinh 2
\left(
\sqrt{H^TH}
\right)
}{2\sqrt{H^TH}}.
\label{stfmf}\end{eqnarray}
Here we put for simplicity $G=1$.
In particular, if $H$ is symmetric, we will have
$$
U=\tilde{U}=\exp(H),\hspace{1em}
g=\exp(2H).
 $$

\subsection{Low-energy limit}
We will call a spacetime function or
a tensor field $\psi$  a slowly varying
field of the order $k$ if it satisfies
$$
\frac{\partial^n}{\partial^n X^{\eta}}\psi=0(\epsilon^{n+k}).
$$
Here $\epsilon$ is an infinitesimally small parameter characterizing
the scale of energy.
If the value of $k$ is not specified, we will assume that
it is trivial.
For a deformation corresponding to a vertex operator function
$\Psi=H_{\nu\mu}\partial X^{\nu}\bar{\partial} X^{\mu}$ with
a slowly varying field $H_{\nu\mu}$, the asymptotic behavior of
the coordinate $T$-product can be shown to be  approximately
the same as for deformation with a constant  field;
{\em i.e.\/},
$$
\langle
X^{\nu}(z^{\prime})X^{\mu}(z)
\rangle_{\Sigma}^{\Psi}
\approx-
2\langle
:\!g^{\nu\mu}(z)\! :
\rangle_{\Sigma}^{\Psi}
\ln|z^{\prime}-z|+ O(\epsilon).
$$
Here
$g^{\nu\mu}$
is a contravariant metric tensor
defined in~(\ref{stfmf}).
In a more covariant way this formula can be written as
\begin{equation}\label{stvg}
\langle
:\!\psi(z_1)\! :\,:\!\phi(z_2)\! :
\rangle_{\Sigma}^{\Psi}
\approx
-2\langle
:\!\psi^{;\nu}\phi_{;\nu}\! :
\rangle_{\Sigma}^{\Psi}
\ln|z^{\prime}-z|+O(\epsilon^3).
\end{equation}
Here $\psi$, $\phi$ are slowly varying space time functions.
As usual,
superscript indices following a semicolon denote derivatives of
spacetime functions
or covariant (Christoffel) derivatives of spacetime tensor fields,
and
the covariant and contravariant metric tensors are used to to raise
and lower
indices.
As a consequence of~(\ref{stvg}) we have
\begin{equation}\label{stvr}
:\!\psi^{;\nu}\phi_{;\nu}(z)\! : = -
{\rm Res}_{z^{\prime}=z}^{\Psi}
\partial^{\Psi}\!:\!\psi(z^{\prime})\!
:\,:\!\phi(z)\! :+O(\epsilon^4).
\end{equation}
Let us take the antiholomorphic derivative of both sides
of this equation
\begin{eqnarray}
\bar{\partial}^{\psi}\!:\!\psi^{;\nu}\phi_{;\nu}(z)\! :
&=&
-{\rm Res}_{z^{\prime}=z}^{\Psi}
\bar{\partial}^{\Psi}\!\partial^{\Psi}\!:\!\psi(z^{\prime})
\! :\, :\!\phi(z)\! :
\nonumber\\[1ex] &&
-{\rm Res}_{z^{\prime}=z}^{\Psi}\partial^{\Psi}:\!\psi(z^{\prime})\!
:\,
\bar{\partial}^{\Psi}:\!\phi(z)\! :
+O(\epsilon^5).
\label{stv1}\end{eqnarray}
In the most general case, the action of the deformed conformal
Laplacian  here in the low-energy limit can be given as
\begin{equation}\label{stvL}
\bar{\partial}^{\Psi}\!\partial^{\Psi}\!:\!\psi\! :=
:\!\left(
\psi_{;\nu\mu}+i\psi^{;\eta}C_{\eta\nu\mu}
\right)
\partial^\Psi\! X^{\nu}\bar{\partial}^\Psi\! X^{\mu}\! :
+O(\epsilon^4) .
\end{equation}
Here  $C_{\eta\nu\mu}$ is the first order slowly varying tensor
field.
This field, if it is not trivial,
violates chiral invariance, and, in some sense,
makes the theory heterotic.
Let us substitute~(\ref{stvL}) to ~(\ref{stv1}),
\begin{equation}
{\rm Res}_{z^{\prime}=z}^{\Psi}
\partial^{\Psi}\!:\!\psi(z^{\prime})\!
:\,\bar{\partial}^{\Psi}\!:\!\phi(z)\!:=
-:\!(\phi_{;\mu\eta}-iC_{\eta\nu\mu}\phi_{;\nu})
\psi^{;\eta}\bar{\partial}^{\Psi}\! X^{\mu}\!:
+O(\epsilon^5)
\label{stvha}
\end{equation}
and then apply~(\ref{fd}) to the right-hand side:
\begin{equation}
{\rm Res}_{\bar{z}^{\prime}=\bar{z}}^{\Psi}
\bar{\partial}^{\Psi}\!:\!\psi(z^{\prime})\! :\,
\bar{\partial}^{\Psi}\!:\!\phi(z)\! :=
-:\!(\phi_{;\mu\eta}-iC_{\eta\nu\mu}\phi_{;\nu})
\psi^{;\eta}\bar{\partial}^{\Psi}\! X^{\mu}\!:
+O(\epsilon^5) .
\label{stvaa}
\end{equation}
The action of the residuelike operations~(\ref{lh}) on vertex
operator
functions with slowly varying coefficients can be given as
\begin{eqnarray}
{\cal A}_{\bar{z}^{\prime}=\bar{z}}^{\Psi}\,\bar{\partial}^{\Psi}
:\!\psi(z^{\prime})\! :\bar{\partial}^{\Psi}
:\!\phi(z)\! :
&=&-:\!\psi^{;\nu}\phi_{;\nu}\!
:+O(\epsilon^4)\nonumber\\[1ex]
{\cal B}_{\bar{z}^{\prime}=\bar{z}}^{\Psi}\,\bar{\partial}^{\Psi}
:\!\psi(z^{\prime})\! :\bar{\partial}^{\Psi}
:\!\phi(z)\! :&=&O(\epsilon^4) .
\label{stvlb}\end{eqnarray}
Substituting it to~(\ref{fdab}) and then applying~(\ref{stvaa}),
we will have
$$
C_{\eta\nu\mu}
\left(
\psi_{;\eta}\phi_{;\nu}+
\phi_{;\eta}\psi_{;\nu}
\right)
= O(\epsilon^5) .
 $$
Therefore,
$$
C_{\eta\nu\mu}+C_{\nu\eta\mu}= O(\epsilon^3).
 $$
Analogously, using conjugated equations it can be shown that
$$
C_{\eta\nu\mu}+C_{\mu\eta\nu}=O(\epsilon^3).
 $$
Thus, $C_{\eta\nu\mu}$ is a completely skew symmetric tensor field.
As a consequence of it,  this field is real for CP-invariant
deformation.
We can generalize
Eqs.~(\ref{stvL}),~(\ref{stvr})~(\ref{stvha})
and ~(\ref{stvaa})
substituting the fields $\partial^{\Psi}\!:\!\psi\! :$,
$\bar{\partial}^{\Psi}\!:\!\phi\! :$
by the vertex operator functions of a more general type
$:\!k_{\nu}\partial^{\Psi}\!X^{\nu}\! :$,
$:\!u_{\nu}\bar{\partial}^{\Psi}\!X^{\nu}\! :$
\begin{eqnarray}
&&\hspace{-1.5em}{\rm Res}_{z^{\prime}=z}:\!k_{\nu}\partial^{\Psi}\!
X^{\nu}(z^{\prime})\! :\,:\!\psi(z)\!:=
-k^{\eta}\psi_{;\eta}+O(\epsilon^3) ,
\label{stvhr}\\[1ex]&&\hspace{-1.5em}
{\rm Res}_{\bar{z}^{\prime}=
\bar{z}}:\!k_{\mu}\bar{\partial}^{\Psi}\!
X^{\mu}(z^{\prime})\! :\,
:\!\psi(z)\! :=
-k^{\eta}\psi_{;\eta}+O(\epsilon^3) ,
\label{stvar}\\[1ex]&&\hspace{-1.5em}
\bar{\partial}^{\Psi}:\!k_{\nu}\partial^{\Psi}\! X^{\nu}\! :
=:\!\left(
k_{\nu;\mu}+ik^{\eta}C_{\eta\nu\mu}
\right)
\partial\!X^{\nu}\bar{\partial}\!X^{\mu}\!:
+O(\epsilon^3) ,
\label{stvlh}\\[1ex]&&\hspace{-1.5em}
\partial^{\Psi}\!:\!k_{\mu}\bar{\partial}^{\Psi}X^{\mu}\!:
=:\!\left(
k_{\mu;\nu}+ik^{\eta}C_{\eta\nu\mu}
\right)
\partial\!X^{\nu}\bar{\partial}\!X^{\mu}\!:
+O(\epsilon^3) ,
\label{stvla}\\[1ex]&&\hspace{-1.5em}
{\rm Res}_{z^{\prime}=z}^{\Psi}
:\!k_{\nu}\partial^{\Psi}\!X^{\nu}(z^{\prime})\!:\,
:\!u_{\mu}\bar{\partial}^{\Psi}X^{\mu}(z)\!:
=-:\!k^{\eta}
\left(
u_{\mu;\eta}-iC_{\eta\nu\mu}u_{\nu}
\right)
\bar{\partial}^{\Psi} X^{\mu}\! :
\nonumber\\&&\hspace{15em}
+O(\epsilon^3),
\label{stvhag}\\[1ex]&&\hspace{-1.5em}
{\rm Res}_{\bar{z}^{\prime}=\bar{z}}^{\Psi}
:\!k_{\eta}\bar{\partial}^{\Psi}X^{\eta}(z^{\prime})\! :  \,
:\!u_{\mu}\bar{\partial}^{\Psi}X^{\mu}(z)\! :
=-:\!\left(
k^{\eta}
\left(
u_{\mu;\eta}-iC_{\eta\nu\mu}u_{\nu}
\right)
+dk_{\eta\mu}U^{\eta}
\right)
\nonumber\\&&\hspace{15em}\times\bar{\partial}^{\Psi} X^{\mu}
\! : +O(\epsilon^3) .
\label{stvaag}\end{eqnarray}
The term in the last equation including external differential
$
dk_{\eta\mu}=k_{\eta;\mu}-k_{\mu;\eta}
 $
is intended to satisfy  formula~(\ref{fdab}).
It is trivial for a gradient tangent field $k_{\nu}=\psi_{;\nu}$.

We can use the formula~(\ref{stvaag}) and
the formula conjugated to~(\ref{stvhag})
to calculate the  residue
\begin{eqnarray}
&&
{\rm Res}_{\bar{z}^{\prime}=\bar{z}}^{\Psi}
:\!k_{\eta}\bar{\partial}^{\Psi}\!X^{\eta}(z^{\prime})\! :\,
:\!m_{\nu\mu}\partial^{\Psi}\!X^{\nu}
\bar{\partial}^{\Psi}\!X^{\mu}(z)\!:
\nonumber\\&&
=-:\!k^{\eta}
\left(
 m_{\nu\mu;\eta}
-iC_{\eta\nu\sigma}m^{\sigma}\!_{\mu}
-iC_{\eta\sigma\mu}m_{\nu}\!^{\sigma}
\right)
\partial^{\Psi}\!X^{\nu}
\bar{\partial}^{\Psi}\!X^{\mu}\!:
\nonumber\\&&\hspace{1em}
-:\!dk_{\eta;\mu}m_{\nu}\!^{\eta}
\partial^{\Psi}\!X^{\nu}
\bar{\partial}^{\Psi}\!X^{\mu}\! :
 +O(\epsilon^3).
\label{stvam}\end{eqnarray}
Applying here~(\ref{lh}) and taking into account that
\begin{eqnarray}
{\cal A}_{\bar{z}^{\prime}=\bar{z}}^{\Psi}
:\!k_{\eta}\bar{\partial}^{\Psi}\!X^{\eta}(z^{\prime})\! :\,
:\!m_{\nu\mu}\partial^{\Psi}\!X^{\nu}\bar{\partial}^{\Psi}X^{\mu}(z)\!
:
&=&
-k^{\mu}m_{\nu\mu}\partial^{\Psi}
X^{\nu}+O(\epsilon^2),
\nonumber\\
{\cal A}_{\bar{z}^{\prime}=\bar{z}}^{\Psi}
:\!k_{\eta}\bar{\partial}^{\Psi}\!X^{\eta}(z^{\prime})\! :\,
:\!m_{\nu\mu}\partial^{\Psi}X^{\nu}\bar{\partial}^{\Psi}\!X^{\mu}(z):
&=&O(\epsilon^2),
\nonumber
\end{eqnarray}
we will also have
\begin{eqnarray}
&&{\rm Res}_{\bar{z}^{\prime}=\bar{z}}^{\Psi}\,
:\!m_{\nu\mu}\partial^{\Psi}\!X^{\nu}
\bar{\partial}^{\Psi}\!X^{\mu}(z^{\prime})\! :\,
:\!k_{\eta}\bar{\partial}^{\Psi}\!X^{\eta}(z)\! :
\nonumber\\
&&=-:\!k^{\eta}
\left(
m_{\nu\eta;\mu}-m_{\nu\mu;\eta}
+iC_{\sigma\nu\mu}m^{\sigma}\!_{\eta}
+iC_{\eta\nu\sigma}m^{\sigma}\!_{\mu}
+iC_{\eta\sigma\mu}m_{\nu}\!^{\sigma}
\right)
\partial^{\Psi}\!X^{\nu}\bar{\partial}^{\Psi}\!X^{\mu}\!:
\nonumber\\ &&\hspace{1em}
-:\!m_{\nu}\!^{\sigma}k_{\mu;\sigma}
\partial^{\Psi}\!X^{\nu}\bar{\partial}^{\Psi}\!X^{\mu}\!:
+O(\epsilon^3).
\label{stvma}\end{eqnarray}
Now, let us calculate a holomorphic derivative of both sides
of~(\ref{stvaa}):
\begin{eqnarray}&&\hspace{-2em}
{\rm Res}_{\bar{z}^{\prime}=\bar{z}}^{\Psi}
:\!\left(
\psi_{;\nu\mu}+i\psi^{;\eta}C_{\eta\nu\mu}
\right)
\partial^{\Psi}\!X^{\nu}
\bar{\partial}^{\Psi}\!X^{\mu}(z^{\prime})\!:\,
\bar{\partial}^{\Psi}\!:\!\phi(z)\! :
\nonumber\\&&\hspace{-2em}
+{\rm Res}_{\bar{z}^{\prime}=\bar{z}}^{\Psi}
\bar{\partial}^{\Psi}\!:\!\psi(z^{\prime})\!:\,
:\!\left(
\phi_{;\nu\mu}+i\phi^{;\sigma}C_{\sigma\nu\mu}
\right)
\partial^{\Psi}\!X^{\nu}
\bar{\partial}^{\Psi}\!X^{\mu}(z^{\prime})\! :
\nonumber\\[1ex]&&\hspace{-2em}
=-:\!\left(
\psi^{;\eta}\phi_{;\eta\nu\mu}+
\psi_{;\nu}\!^{\eta}\phi_{;\mu\eta}+
C_{\rho\nu\mu}C^{\rho\eta\sigma}\psi_{;\eta}\phi_{;\sigma}
\right)
\partial^{\Psi}\!X^{\nu}
\bar{\partial}^{\Psi}\!X^{\mu}\! :
\nonumber\\&&\hspace{-1em}
-i:\!\left(
C_{\sigma\nu\mu}\psi_{;\eta}\phi^{;\sigma\eta}
-C_{\eta\sigma\mu}
\left(
\psi^{;\eta}\phi^{;\sigma}
\right)_{;\nu}
-C_{\eta\sigma\mu;\nu}\psi^{;\eta}\phi^{;\sigma}
\right)
\partial^{\Psi}\!X^{\nu}
\bar{\partial}^{\Psi}\!X^{\mu}\! :
+O(\epsilon^6).
\nonumber\end{eqnarray}
Applying here~(\ref{stvma})
and~(\ref{stvam}), we will have
\begin{equation}\label{stvCc}
C_{\sigma\nu\mu;\eta}+C_{\eta\nu\sigma;\mu}
-C_{\eta\nu\mu;\sigma}+C_{\eta\sigma\mu;\nu}=O(\epsilon^4)
\end{equation}
or, in a geometric form,
$$
dC=O(\epsilon^4).
 $$
This means that $C_{\eta\nu\mu}$ is a cocycle differential
three-form.
If topology of the spacetime is trivial, all cocycles are exact and,
therefore,
the field  $C_{\eta\nu\mu}$
can be  represented as
\begin{equation}\label{stvco}
C=d\omega + O(\epsilon^3).
\end{equation}
Here $\omega$ is a slowly varying differential two-form;  {\em
i.e.\/},
a skew symmetric rank 2 tensor field.
In order to express $\omega_{\nu\mu}$  through the
deformation parameters $H_{\nu\mu}$,
we will again consider a family of deformation theories
corresponding to scaled fields $\tau H_{\nu\mu}$ and calculate
a derivative of the deformed Laplacian~(\ref{stvL})
with respect to $\tau$
\begin{eqnarray}
&&\hspace{-2em}
\frac{\partial}{\partial\tau}\!\partial^{\tau\Psi}
\!\bar{\partial}^{\tau\Psi}\!\psi=
:\!\left(
\frac{\partial}{\partial\tau}\psi_{;\nu\mu}+
i\psi_{;\eta}C_{\sigma\nu\mu}
\frac{\partial}{\partial\tau}g^{\eta\sigma}
+i\psi^{;\eta}\frac{\partial}{\partial\tau}\!C_{\eta\nu\mu}
\right)
\partial^{\tau\Psi}\!X^{\nu}\bar{\partial}^{\tau\Psi}\!X^{\mu}\! :
\nonumber\\&&\hspace{-2em}
+(\psi_{;\nu\mu}+i\psi^{;\eta}C_{\eta\nu\mu})
\left(
\bar{\partial}^{\tau\Psi}X^{\mu}
\frac{\partial}{\partial\tau}\partial^{\tau\Psi}X^{\nu}
+\partial^{\tau\Psi}X^{\nu}\frac{\partial}{\partial\tau}
\bar{\partial}^{\tau\Psi}X^{\mu}
\right) .
\label{stve}\end{eqnarray}
Using formulas~(\ref{dhd}) and~(\ref{stvma})
on the right-hand side of this equation, we will have
\begin{eqnarray}
&&\hspace{-4em}
\frac{\partial}{\partial\tau}\!\partial^{\tau\Psi}\!
\bar{\partial}^{\tau\Psi}\psi=
\nonumber\\[1ex]&&\hspace{-4em}
\partial^{\tau\psi}
\left(
h_{\eta\mu}\psi^{;\eta}\bar{\partial}^{\tau\psi}X^{\mu}
\right)
-{\rm Res}_{\bar{z}^{\prime}=\bar{z}}^{\Psi}\,
:\!h_{\nu\mu}\partial^{\tau\Psi}\!X^{\nu}
\bar{\partial}^{\tau\Psi}\!X^{\mu}(z^{\prime})\!:\,
:\!\psi_{;\eta}\bar{\partial}^{\Psi}\!X^{\eta}(z)\!:
\nonumber\\[1ex]&&\hspace{-4em}=
:\!\left(
\left(
\psi_{;\nu\sigma}+
i\psi^{;\eta}C_{\eta\nu\sigma}
\right)
h^{\sigma}\!_{\mu}+
\left(
\psi_{;\sigma\mu}+
i\psi^{;\eta}C_{\eta\sigma\mu}
\right)
h_{\nu}\!^{\sigma}\,\!:
\partial^{\Psi}\!X^{\nu}
\bar{\partial}^{\Psi}\!X^{\mu}
\right)
\nonumber\\&&\hspace{-3em}+
:\!\left(
\psi^{;\eta}
\left(
h_{\nu\eta;\mu}-h_{\nu\mu;\eta}+
h_{\eta\mu;\nu}
\right)
+2iC_{\sigma\nu\mu}b^{\sigma}\!_{\eta}
\right)
\partial^{\Psi}\!X^{\nu}\bar{\partial}^{\Psi}\!X^{\mu}\!:
+O(\epsilon^4).
\label{stvlc}\end{eqnarray}
Here  $b_{\nu\mu}$ and $ia_{\nu\mu}$ are, respectively, symmetric and
skew symmetric parts
of $h_{\nu\mu}$; {\em i.e.\/},
$$
h_{\nu\mu}=b_{\nu\mu} + ia_{\nu\mu},\hspace{1em}
b_{\nu\mu}=b_{\mu\nu},\hspace{1em}
a_{\nu\mu}=-a_{\mu\nu}.
 $$
According to~(\ref{stfim}), the increment of the metric can be given
as
\begin{equation}\label{stvdg}
\frac{\partial}{\partial\tau}\!g^{\nu\mu}=2b^{\nu\mu},\hspace{1em}
\frac{\partial}{\partial\tau}\!g_{\nu\mu}=-2b_{\nu\mu}.
\end{equation}
The corresponding increments of the Christoffel symbols are
$$
\frac{\partial}{\partial\tau}\!\Gamma_{\nu\mu}^{\eta}=-
\left(
b^{\eta}\!_{\nu;\mu}+b^{\eta}\!_{\mu;\nu}-b_{\nu\mu;\eta}
\right)
 $$
and, therefore,
\begin{equation}\label{stvC}
\frac{\partial}{\partial\tau}\!\psi_{;\nu\mu}=\psi^{;\eta}
\left(
b_{\eta\nu;\mu}+b_{\eta\mu;\nu}-b_{\nu\mu;\eta}
\right) .
\end{equation}
Substituting~(\ref{stvlc}),~(\ref{stvdg}),~(\ref{stvC})
to~(\ref{stve}),
we will come to the following formula for the increment of
$C$:
$$
\frac{\partial}{\partial\tau}\!C_{\eta\nu\mu}=a_{\nu\eta;\mu}
-a_{\nu\mu;\eta}+a_{\eta\mu;\nu}
+O(\epsilon^3),
$$
or, in a geometric  form,
$$
\frac{\partial}{\partial\tau}\!C=-da+O(\epsilon^3).
 $$
This corresponds to the following increment of the field $\omega$:
\begin{equation}\label{stvoe}
\frac{\partial}{\partial\tau}\omega=-a=-\frac{1}{2i}(h-h^T) .
\end{equation}
Integrating this equation we will finally have
$$
\omega=-\frac{1}{2i}\int_{0}^{1}(h-h^T)\, d\!\tau .
 $$
According to~(\ref{stfh}) and~(\ref{stfde}),
$h(\tau)$
can be calculated in one of the following ways:
$$
h=(U^{-1})^T H \tilde{U}^{-1},
-\frac{\partial}{\partial\tau}\!U^{-1}(U^{-1})^T
=
-(\tilde{U}^{-1})^T\frac{\partial}{\partial\tau}\!\tilde{U}^{-1} .
 $$
Local space-time transformations
\begin{equation}
\delta\!g_{\nu\mu}=\varepsilon_{\nu;\mu}
+\varepsilon_{\mu;\nu},
\hspace{1em}
\delta\!\omega_{\nu\mu}=
\varepsilon^{\eta}\omega_{\nu\mu;\eta}+
\varepsilon^{\nu;\eta}\omega_{\eta\mu}
+\varepsilon^{\mu;\eta}\omega_{\nu\eta},
\label{Enstine}
\end{equation}
change the vertex-operator relations
established above to their equivalents.
In addition, these relations will not change at all
under transformation
\begin{equation}\label{DeRam}
\delta{\omega}=d\varsigma.
\end{equation}
Here $\varsigma$ is a differential one-form;  {\em i.e.\/},
cotangent field.
The transformations~(\ref{Enstine})
and~(\ref{DeRam}) are a particular
case of more
general symmetries~(\ref{sym}) with the following
choice of the vertex operator functions parametrizing them:
$$
\xi_z=
:\!(\varepsilon_{\nu}+i\varsigma_{\nu})\partial\!X^{\nu}\!:,
\hspace{1em}
\xi_{\bar{z}}=
:\!(\varepsilon_{\mu}-i\varsigma_{\mu})
\bar{\partial}\!X^{\mu}\!: .
 $$

In string theory, the off-shell theory and symmetries depend,
as we have seen,
on regularization of the contact divergence.
Therefore, changing regularization
parameters $\Lambda(\alpha)$ we will also change the symmetry
algebra.
Not all such algebras can be closed because the Lie algebraic
structure is usually rigid. Of course,
one can try to find a specific
regularization method for which the symmetry algebra is closed.
However, there is no indication that it can be done.
In the low-energy limit the picture is almost independent of
regularization,
which actually allowed the symmetry algebra to be closed.

\subsection{Equation of motion}
Let us consider an action of the deformed Virasoro operator
\begin{equation}\label{stlr}
L_1^{\Psi}={\rm Res}_{z^{\prime}=z}^{\Psi}(z^{\prime}-z)^2
T_{zz}^{\Psi}(z^{\prime})
\end{equation}
on the vertex $:\!k_{\nu}\partial^{\Psi} X^{\nu}\! :$.
In the locally Galilean system of coordinates we can use for
$T_{zz}^{\Psi}$
the formula
$$
T_{zz}^{\Psi}=\frac{1}{2}{\lim_{z^{\prime}\rightarrow z}}^{\Psi}
\!\left(
\partial^{\Psi} X^{\eta}(z^{\prime})\partial^{\Psi}X_{\eta}(z)+
\frac{D}{(z^{\prime}-z)^2}
\right)
+O(\epsilon^2).
 $$
Substituting it to~(\ref{stlr}) and
applying~(\ref{stvhr}) and~(\ref{stvhag}) we will have
\begin{eqnarray}
L_1^{\Psi}:\!k_{\nu}\partial^{\Psi}\!X^{\nu}\!:&=&
-{\rm Res}_{z^{\prime}=z}^{\Psi}
\partial^{\Psi}\!X^{\nu}(z^{\prime})
:\!k_{\nu}(z)\!:
=:\!k_{\nu}\!^{;\nu}\!:
+O(\epsilon^3),
\nonumber\\[1ex]
L_1^{\Psi}:\!f_{\nu\mu}\partial^{\Psi}\!X^{\nu}
\bar{\partial}^{\Psi}\!X^{\mu}\!:&=&
-{\rm Res}_{z^{\prime}=z}^{\Psi}
\partial^{\Psi}\!X^{\nu}(z^{\prime})
:\!f_{\nu\mu}\bar{\partial}^{\Psi}\!X^{\mu}(z)\!:
\nonumber\\&=&
:\!(
f_{\nu\mu}\!^{;\nu}-if^{\nu\eta}C_{\nu\eta\mu})
\bar{\partial}^{\Psi}X^{\mu}\!:
+O(\epsilon^3).
\nonumber\end{eqnarray}
Using this formula together with~(\ref{stvla}) one can show that
\begin{eqnarray}
\bar{\partial}^{\Psi}\!L_1^{\Psi}\,
:\!k_{\nu}\partial^{\Psi}\!X^{\nu}\!:&=&
:\!k^{\eta}\!_{;\eta\mu}\bar{\partial}^{\Psi}\!X^{\mu}\!:
+O(\epsilon^4)\nonumber\\[1ex]
L_1^{\Psi}\bar{\partial}^{\Psi}\,
:\!k_{\nu}\partial^{\Psi} X^{\nu}\!:&=&
:\!\left(
k^{\mu}\!_{;\nu\mu}+
k^{\nu}
(C_{\nu}\!^{\sigma\rho}C_{\mu\sigma\rho}
-iC_{\eta\nu\mu}\!^{;\eta})
\right)
\bar{\partial}^{\Psi}\!X^{\mu}\! :
+O(\epsilon^4),
\nonumber\end{eqnarray}
and, therefore,
\begin{equation}
\left[
\bar{\partial}^{\Psi}, L_1^{\Psi}
\right]
:\!k^{\nu}\bar{\partial}^{\Psi} X^{\nu}\! :
=:\!\left(
R_{\nu\mu}-C_{\nu}\!^{\sigma\rho}C_{\mu\sigma\rho}+
iC_{\eta\nu\mu}\!^{;\eta}
\right)
k^{\nu}\bar{\partial}^{\Psi}\!X^{\mu}\! :
+O(\epsilon^4).
\label{stdc}\end{equation}
Here we use the formula
\begin{equation}\label{stdrc}
k^{\nu}\!_{;\nu\mu}-k^{\nu}\!_{;\mu\nu}=R_{\nu\mu}k^{\nu},
\end{equation}
where $R_{\nu\mu}$ is the Ricci curvature.
Taking antiholomorphic derivatives of~(\ref{stlr})
and applying consequently formulas~(\ref{cl})
and~(\ref{fd}),
we can express the commutator
$\left[
\bar{\partial}^{\Psi}, L_1^{\Psi}
\right]$
through the contradiagonal components of the energy-momentum tensor:
\begin{eqnarray}
\left[
\bar{\partial}^{\Psi}, L_1^{\Psi}
\right]&=&
{\rm Res}_{z^{\prime}=z}^{\Psi}(z^{\prime}-z)^2
\bar{\partial}^{\Psi}\!T_{zz}^{\Psi}(z^{\prime})
-{\rm Res}_{z^{\prime}=z}^{\Psi}
(z^{\prime}-z)^2
\partial^{\Psi}\!T_{z\bar{z}}^{\Psi}(z^{\prime})
\nonumber\\[1ex]&=&
2{\rm Res}_{z^{\prime}=z}^{\Psi}(z^{\prime}-z)
\partial^{\Psi}\!T_{z\bar{z}}^{\Psi}(z^{\prime})
-{\rm Res}_{z^{\prime}=z}^{\Psi}
\partial_{z^{\prime}}^{\Psi}\!\left(
(z^{\prime}-z)^2
T_{z\bar{z}}^{\Psi}(z^{\prime})
\right)
\nonumber\\[1ex]&=&
2{\rm Res}_{z^{\prime}=z}^{\Psi}(z^{\prime}-z)\partial^{\Psi}
T_{z\bar{z}}^{\Psi}(z^{\prime})
-{\rm Res}_{z^{\prime}=z}^{\Psi}(z^{\prime}-z)^2\bar{\partial}^{\Psi}
T_{z\bar{z}}^{\Psi}(z^{\prime}) .
\nonumber\end{eqnarray}
Comparing this with~(\ref{stdc}) we will have
\begin{eqnarray}
T_{z\bar{z}}^{\Psi}&=&
-\frac{1}{2}
:\!\left(
R_{\nu\mu}-C_{\nu}\!^{\sigma\rho}C_{\mu\sigma\rho}+
iC_{\eta\nu\mu}\!^{;\eta}
\right)
\partial^{\Psi}\!X^{\nu}
\bar{\partial}^{\Psi}\!X^{\mu}\!:
+O(\epsilon^4).
\label{stdem}
\end{eqnarray}
Analogously, we can derive the same formula for
$T_{\bar{z} z}^{\Psi}$;  {\em i.e.\/},
\begin{equation}\label{stdlr}
T_{\bar{z} z}^{\Psi}=T_{z\bar{z}}^{\Psi}+O(\epsilon^4) .
\end{equation}
Therefore, the equations of motion~(\ref{eqg})  can be written as
\begin{equation}\label{stde}
T_{z\bar{z}}^{\Psi}=-
\bar{\partial}^{\Psi}\Phi_{z}=-
\partial^{\Psi}\Phi_{\bar{z}} + O(\epsilon^4).
\end{equation}
If the background fields have no symmetries, this is
equivalent to
$$
T_{z\bar{z}}^{\Psi}+\partial^{\Psi}\bar{\partial}^{\Psi}\Phi =
O(\epsilon^4).
 $$
Here $\Phi$ is some vertex operator function.
In order to make this equation solvable with $T_{z\bar{z}}^{\Psi}$
given by formula~(\ref{stdem}) we shall put
$$
\Phi=:\!\phi\! :,
 $$
where $\phi$ is some space-time function.
Then we will come to the following
equation of motion in the low-energy limit
$$
R_{\nu\mu}+iC_{\eta\nu\mu}\!^{;\eta}=
C_{\nu}\!^{\sigma\rho}C_{\mu\sigma\rho}+
2\left(
\phi_{;\nu\mu}+i\phi^{;\eta}C_{\eta\nu\mu}
\right) .
 $$
This equation decouples to symmetric and skew symmetric parts
\begin{eqnarray}
R_{\nu\mu}&=&C_{\nu}\!^{\sigma\rho}C_{\mu\sigma\rho}+2\phi_{;\nu\mu},
\nonumber\\[1ex]
C_{\eta\nu\mu}\!^{;\eta}&=&2\phi^{;\eta}C_{\eta\nu\mu}.
\label{std!}\end{eqnarray}
The skew symmetric part can be also written as
$$
\left(
\frac{1}{\Theta}C_{\eta\nu\mu}
\right)_{;\eta}=0,
 $$
or, in a geometric form, as
$$
d\left(
\frac{1}{\Theta}C^{\ast}
\right)=0.
 $$
Here
$\Theta=\exp2\phi$ is a so-called {\em dilaton} field.
If the field  $C_{\eta\nu\mu}$ is trivial, the first equation
in~(\ref{std!})
will be a conventional Brans-Dicke equation
derived earlier in the  $\beta$-function approach.
Note that the  transformations
$$
\phi\ \longrightarrow \ \phi +{\rm const}
 $$
do not change the corresponding CFT and, therefore, only the
derivatives of $\phi$ are physically
important.
These derivatives must be bounded and slowly varying. However, the
function
$\phi$ itself should not necessarily comply with either of the
conditions.

\subsection{Deformation of central charge}
Let us differentiate the first equation in~(\ref{std!}) and
then contract indices
$$
R_{\nu\mu}\!^{;\nu}=C^{\nu\sigma\rho}\!_{;\nu}C_{\mu\sigma\rho}+
C^{\nu\sigma\rho}C_{\mu\sigma\rho;\nu}+2\phi_{;\nu\mu}\!^{\nu}.
 $$
Using here the formulas ~(\ref{stvCc}) and~(\ref{stdrc})
and  the identity
 $$
R_{\nu\mu}\!^{;\nu}=\frac{1}{2}R_{;\mu}\hspace{1em}
(R\equiv R^{\nu}\!_{\nu}),
 $$
respectively, on the right and left hand sides we will have
$$
\frac{1}{2}R_{;\mu}=-4\phi_{;\nu\mu}\phi_{\nu}+
\frac{1}{3}C^{\nu\sigma\rho}C_{\nu\sigma\rho;\mu}.
 $$
This equation can be easily integrated
\begin{equation}\label{stci}
R+4\phi^{;\nu}\phi_{;\nu}-
\frac{1}{3}C^{\nu\sigma\rho}C_{\nu\sigma\rho}=
{\rm const}\equiv m^2.
\end{equation}
Contracting indices in the first equation in~(\ref{std!}) we will
also have
$$
R=C^{\nu\sigma\rho}C_{\nu\sigma\rho}+
2\Box\phi.
 $$
Using this formula we can exclude the curvature from the~(\ref{stci})
$$
2\Box\phi + 4\phi^{;\nu}\phi_{;\nu}=m^2-
\frac{2}{3}C^{\nu\sigma\rho}C_{\nu\sigma\rho}.
 $$
This is equivalent to the following  equation on the dilaton field
$$
\Box\Theta=\left(
m^2-\frac{2}{3}C^{\nu\sigma\rho}C_{\nu\sigma\rho}
\right)
\Theta.
 $$
Therefore, $m$ can be interpreted as a dilaton mass.
It is the only topological characteristic of the space-time
dynamics.
Therefore, it must be related to the only topological characteristic
of CFT -- a central charge.
We can easily find this relation using the  flat solution
$$
g_{\nu\mu}={\rm const},\hspace{1em} C_{\eta\nu\mu}=0,\hspace{1em}
\phi=k_{\nu}x^{\nu},\hspace{1em}
m^2=4k^{\nu}k_{\nu},
 $$
corresponding to to the trivial parameter $\Psi$.
The deformed energy-momentum
tensor~(\ref{emc}), in this case, can be given as
$$
T_{zz}^{\Phi}=T_{zz}-\partial\Phi_z=
-\frac{1}{2}
:\!\partial X^{\nu}\partial X_{\nu}\! :
-k_{\nu}
(\partial)^2{\bf X}^{\nu}.
 $$
Its two-point correlation function has
the contact singularity
$$
\langle
T_{zz}^{\Phi}(z_1)T_{zz}^{\Phi}(z_2)
\rangle_{\Sigma}
\approx\frac{\frac{D}{2}+
\frac{1}{4}m^2}{(z_1-z_2)^4}{\langle 0
\rangle}_{\Sigma} .
 $$
This corresponds to the central charge equal to
\begin{equation}\label{stdcd}
c= D+\frac{1}{2}m^2.
\end{equation}
In particular, the conventional Brans-Dicke theory with the massless
dilaton describe critical deformations of CFT.

\sect{Problems and Perspectives}
In this paper we formulated the equation of motion for
the closed string field theory and described its symmetries.
In order to do this we introduced
a specific regularization of the
contact singularities of the $T$-product of vertex operators.
Despite the fact that classes of gauge equivalent solutions
correspond to
equivalence classes of CFT and, therefore, do not depend on
this regularization, the off-shell states,
the equation of motion and the symmetries depend on it.
In addition, the symmetries as well as equivalence relations between
solutions in different regularizations are very nontransparent.
It makes it difficult to formulate a classical action corresponding
to the equation we found here and, therefore, to quantize the
theory.  Some ideas on how this problem may be solved by means of
auxiliary fields were suggested in~\cite{egn}.
However, we are not sure that the string theory
can be, in principle, covariantly quantized
in the canonical path integral approach,
which requires the action to be covariantly defined,
and hope that that there is a  more explicit approach to describe
the quantum string field theory.

Solving the equation of motion derived here we may
try to find  string vacuum which is crucial
for obtaining  results
observable in the low-energy experiments.
Because of the infinite number
of the higher order terms
in this  equation, it is not an easy task.

The formalism suggested here works
equally well for the string theory with all possible central charges.
 Moreover, it gives a mechanism for central charge
deformation~(\ref{stdcd}).
As a negative consequence of this, the massive symmetries
corresponding to critical string are not understood.
These symmetries may not be  related to
equivalence relations of CFT. It is a very important principle
drawback of interpretation of the closed string field theory
as the  theory of CFT's.
More recently we have found  that CFT induces certain
noncommutative functor
from the category of two dimensional surfaces and suggest
that such functors, rather then CFT's, must be associated with
classical string states.
This might explain the origin of the additional critical symmetries
and help to close symmetry algebra.
It is especially encouraging
to see that deformations of such functor can be
associated with certain $BRST$-like cohomologies.
\sect{Acknowledgements}
I am very grateful to Mark Evans for inspiring me to develop the
{\em deformation} approach, for discussions and for many
important suggestions.
It is also a pleasure to thank Toni Weil for significant editing
help.
\newpage\appendix
\sect{Successor Calculation}
\label{appx}
Here we will present a method for calculation of successors
defined in Section~\ref{drlo}.
In order to use the formula~(\ref{sc1})
we should first learn how to calculate $\partial_{\bar{z}}^{-1}F$ for
an arbitrary smooth function $F$, having diagonal singularities.
Such a function can  always be considered
as a linear combination  of
the following samples:
\begin{equation}\label{sm}
\prod_{i=0}^n |z-z_i|^{-2\alpha_i}\hspace{1em}(\alpha_i\in{\bf R})
\end{equation}
with meromorphic coefficients.
Therefore, the problem can be reduced to calculation of
$\partial_{\bar{z}}^{-1}$
for these samples. Let us use for these samples
the integral representation
\begin{equation}\label{al}
|z-z_i|^{-2\alpha_i}=\frac{1}{\Gamma(\alpha_1)}\int_0^{\infty}
\exp\left(
-s_i|z-z_i|^2
\right)
{s_i}^{\alpha_i-1}\,d\!s_i .
\end{equation}
Then~(\ref{sm}) will be  expressed through integrals over $d\!s_i$ of
the
functions
$$
E_{s_0,\ldots,s_n}=
\exp\left(
-\sum_0^n s_i|z-z_i|^2
\right) ,
 $$
which can be explicitly integrated over $\bar{z}$
\begin{eqnarray}
\partial_{\bar{z}}^{-1}E_{s_0,\ldots,s_n}&=&
\left(
\sum_{k=0}^n s_k(z-z_k)
\right)^{-1}
\left(
\exp\left(
-\sum_0^n s_i|z-z_i|^2
\right)
\right. \nonumber\\
&&\left.
-\exp\left(
-\left(
\sum_{k=0}^n s_k
\right)^{-1}
\sum_{0\leq i \leq j}s_i s_j|z_i-z_j|^2
\right)
\right).
\label{ai}\end{eqnarray}
The second holomorphic term here is meant to cancel the singularity
at
$z=\left(
\sum_{k=0}^n s_k
\right)^{-1}
\sum_{k=0}^n s_k z_k$.
Using this formula we found an explicit result for successors of
rank 2 residuelike operations. For example,
\begin{eqnarray}
&&\hspace{-2em}{\rm Res}_{\bar{z}\doteq\bar{z}_1 =\bar{z}_0}
|z-z_1|^{-2\alpha}|z-z_0|^{-2\beta}
 |z_1-z_0|^{-2\gamma}F
\nonumber\\
&&\hspace{-2em}=
\sum_{\tilde{\imath}+\tilde{\jmath}=i+j+1}\!
\Biggl(
\frac{
\Gamma(\alpha+\beta-\tilde{\imath})\Gamma(i+1-\beta)\Gamma(1-\alpha)
}{
\Gamma(\alpha)\Gamma(\beta-\tilde{\imath})\Gamma(i+2-\alpha-\beta)}
\Lambda(\tilde{\imath}+
\tilde{\jmath}+1-\alpha-\beta-\gamma)
\Biggr.\nonumber\\[1ex]
&&\hspace{-2em}
-\sum_k\left(
\frac{
\Lambda(k+1-\alpha)
}{
k+1-\alpha
}
C_{i-\beta}^kC_{\tilde{\imath}-\beta}^k\Lambda
(\tilde{\imath}+\tilde{\jmath}-k-\beta-\gamma)
\right.\nonumber\\[1ex]
&&\hspace{-2em}
+\frac{
\Lambda(k+1-\beta)
}{
k+1-\beta
}
C_{-\alpha}^{k-i}
C_{-\alpha}^{k-\tilde{\imath}}(-)^{\tilde{\imath}-i}
\Lambda(\tilde{\imath}+\tilde{\jmath}-k-\alpha-\gamma)
\nonumber\\[1ex]
&&\hspace{-2em}
\Biggl.\left.+
\frac{
\Lambda(k+1-\alpha-\beta)
}{k+1-\alpha-\beta}
C_{-\alpha}^{k+i} C_{-\alpha}^{k+\tilde{\imath}}\Lambda
(\tilde{\imath}+\tilde{\jmath}+k-\gamma)
\right)
\Biggr)
\frac{
\partial_{z}^i\partial_{\bar{z}}^{\tilde{\imath}}
\partial_{z_1}^j\partial_{z_1}^{\tilde{\jmath}}F
}{
i!\tilde{\imath}!j\tilde{\jmath}!
}
\nolinebreak\begin{picture}(33,7)(0,0)
\setlength{\unitlength}{.2em}\put(1,-7){\line(0,1){14}}
\put(2,-7){$\scriptstyle z=z_1=z_0$}
\end{picture}.
\label{Hm}\end{eqnarray}
Here $F$ is a function of $z,\ z_1,\ z_0$ regular in some environment
of the
main diagonal $z=z_1=z_0$. The generalized number of composition is
defined to
be
$$
C_{\alpha}^k\equiv
\left\{
\begin{array}{ll}
\frac{\alpha(\alpha-1)\cdots(\alpha-k)}{k!} & (k\geq 0)\\
0 & (k < 0).
\end{array}
\right.
 $$
The poles on the right-hand side of~(\ref{Hm})
completely cancel each other, so that the successor  depends on
parameters $\alpha$, $\beta$ and $\gamma$ regularly.

The formula for this successor can be simplified if it acts on the
function, which
depends on $z_1$ holomorphically. Such a function can be always
represented as
$$
F=\sum_{k\in{\bf Z}}A_k(z,z_0)(z_1-z)^{-1-k}+B_k(z_1-z_0)^{-1-k}
 $$
and, therefore, its antiderivative can be given as
$$
\partial_{\bar{z}}^{-1}F=
\sum_{k\in{\bf Z}}
\partial_{\bar{z}}^{-1}A_k(z,z_0)(z_1-z)^{-1-k}
+\partial_{\bar{z}}^{-1}B_k(z,z_0)(z_1-z_0)^{-1-k}.
 $$
Substituting this to~(\ref{sc1}), and using~(\ref{fd}) we will have
\begin{eqnarray}
{\rm Res}_{z\doteq z_1=z_0}F&=&
{\rm Res}_{z=z_0}\partial_{\bar{z}}^{-1}B_0(z,z_0)
-{\rm Res}_{z=z_0}\partial_{\bar{z}}^{-1}A_0(z,z_0)
-{\rm Res}_{z=z_0}\partial_{\bar{z}}^{-1}B_0(z,z_0)
\nonumber\\&&
+\sum_{k=0}^{\infty}
\frac{(-)^k}{k!}{\rm Res}_{z_1=z_0}
\partial_{z_1}^k\partial_{z_1}^{-1}A_k
\nonumber\\[1ex]&=&
{\rm Res}_{z=z_0}\sum_{k=1}^{\infty}
\frac{(-)^k}{k!}\partial_{z}^k\partial_{\bar{z}}^{-1}A_k
={\rm Res}_{\bar{z}=\bar{z}_0}
\sum_{k=1}^{\infty}\frac{(-)^k}{k!}
\partial_{z}^{k-1}A_k .
\nonumber\end{eqnarray}
Noticing here, that for positive $k$
$$
A_k(z,z_0)={\rm Res}_{z_1=z}(z_1-z)^kF(z,z_1,z_0) ,
 $$
we will come to the following formula for the successor
\begin{equation}
{\rm Res}_{z\doteq z_1=z_0}F=
{\rm Res}_{\bar{z}=\bar{z}_0}
\sum_{k=1}^{\infty}\frac{(-)^k}{k!}\partial_{z}^{k-1}
{\rm Res}_{z_1=z}(z_1-z)^k
F(z,z_1,z_0).
\label{sch}\end{equation}
In this way, we can always express the higher rank residue successor
applied to functions meromorphic with respect to one of the
variables
through the lower rank residue successors.

The calculation of successors can be also  based
on their analytical properties.
Let $\Sigma$ in~(\ref{sc}) be a Riemann sphere $\bar{{\bf C}}$.
The sample functions~(\ref{sm}) may have singularities at $z=\infty$.
For regularization of corresponding divergence of the integral we can
use
the  formula
$$
\int_{\bar{{\bf C}}}=\int_0^{\infty}
d\!\mu(r)\int_{D(0,\frac{1}{r})}.
 $$
Difference between the regularizations corresponding to parameters
$\Lambda$ and $\Lambda^{\prime}$ can be given as
$$
\int_{\bar{{\bf C}}}^{\prime}-\int_{\bar{{\bf C}}}=\sum_{i=0}^n
I_{z\stackrel{\xi}{=}z_i}
+I_{z\stackrel{\xi}{=}\infty}
\hspace{1em}
\left(
\xi(\alpha)=\frac{\Lambda(\alpha)-\Lambda^{\prime}(\alpha)}{\alpha}
\right).
 $$
Here $I_{z\stackrel{\xi}{=}z_0}$
is a residuelike operation defined as
$$
I_{z\stackrel{\xi}{=}z_0}|z-z_0|^{2(\alpha-1)}z^k=
\xi(\alpha)\delta_{k,0},
\hspace{1em}
I_{z\stackrel{\xi}{=}\infty}|z-z_0|^{2(\alpha-1)}z^k
=\xi(-\alpha)\delta_{k,0}.
 $$
Therefore, the combination
$$
{\cal I}(F)=
\int_{\bar{{\bf C}}}F-\sum_{i=0}^n I_{z\stackrel{\xi}{=}z_i}F
-I_{z\stackrel{\xi}{=}\infty}F\hspace{1em}
\left(
\xi=-\frac{\Lambda}{\alpha}
\right)
 $$
does not depend on $\Lambda$. In fact, it
is the analytical continuation of the nonregularized integral from
the divergence-free area of the parameters.
All the poles of ${\cal I}$  are determined
by the singularities of the residuelike operations
resulting from singular behavior of $\xi(\alpha)$ at $\alpha=0$.

\end{document}